\documentclass[conference]{IEEEtran}
\IEEEoverridecommandlockouts
\usepackage{cite}
\usepackage{amsmath,amssymb,amsfonts}
\usepackage{graphicx}
\usepackage{textcomp}
\usepackage{xcolor}
\usepackage{makecell}
\usepackage{tabularx}
\usepackage{algorithm}
\usepackage{algpseudocode}
\usepackage{array}
\usepackage{amsmath}

\usepackage{subcaption}

\usepackage{xspace}
\usepackage{placeins}
\usepackage{float}     
\usepackage{multirow}
\usepackage{flushend}
\usepackage{url}

\usepackage{tikz}
\usetikzlibrary{positioning,fit,arrows.meta}

\def\BibTeX{{\rm B\kern-.05em{\sc i\kern-.025em b}\kern-.08em
    T\kern-.1667em\lower.7ex\hbox{E}\kern-.125emX}}

\begin{document}



\title{
Multi-Layer Scheduling for MoE-Based LLM Reasoning
}

\author{

\IEEEauthorblockN{Yifan Sun\IEEEauthorrefmark{1},
Gholamreza Haffari\IEEEauthorrefmark{2},
Minxian Xu\IEEEauthorrefmark{3},
Rajkumar Buyya\IEEEauthorrefmark{4},
Adel N. Toosi\IEEEauthorrefmark{1}
}
\IEEEauthorblockA{\IEEEauthorrefmark{1}DisNet Lab, School of Computing and Information Systems, The University of Melbourne, Melbourne, Australia}
\IEEEauthorblockA{\IEEEauthorrefmark{2}Faculty of Information Technology, Monash University, Melbourne, Australia}
\IEEEauthorblockA{\IEEEauthorrefmark{3}Shenzhen Institute of Advanced Technology, Chinese Academy of Sciences, Shenzhen, China}
\IEEEauthorblockA{\IEEEauthorrefmark{4}School of Computing and Information Systems, The University of Melbourne, Melbourne, Australia}


}

\maketitle

\begin{abstract}
Large Language Models (LLMs) have achieved remarkable success across a wide range of tasks, but serving them efficiently at scale remains a critical challenge due to their substantial computational and latency demands. While most existing inference frameworks rely on simple scheduling strategies such as First-Come-First-Serve (FCFS) at the engine level and Round-Robin (RR) at the scheduler or coordinator level, they often fail to fully utilize system resources and may suffer from issues such as head-of-line blocking and load imbalance. Recent advances in Mixture-of-Experts (MoE) models have also introduced new challenges in scheduling arising from expert parallelism and routing complexity. This research proposes a multi-layer scheduling framework tailored for MoE-based LLM serving. It targets scheduling at three levels: request-level, engine-level, and expert-level. At the request level, we explore algorithms such as Shortest-Job-First (SJF) and priority-aware aging to improve throughput and reduce latency. At the engine level, we design load-aware dispatching strategies that account for the current prefix token load, KV cache utilization, and user stickiness to achieve better resource matching. At the expert level, we focus on alleviating expert hotspots and strategically placing inter-layer expert dependencies to balance load and improve routing efficiency. Extensive experimental results from more than 100 experiments conducted under diverse workload distributions show that our approach consistently outperforms the state-of-the-art inference framework vLLM, achieving up to 17.8\% reduction in Time To First Token (TTFT) latency and 13.3\% reduction in Time-Per-Output-Token (TPOT) latency.
\end{abstract}

\begin{IEEEkeywords}
AI/ML Driven Distributed Systems
\end{IEEEkeywords}

\section{Introduction}

In recent years, Large Language Models (LLMs) have developed rapidly~\cite{openai2022chatgpt}, helping millions of users worldwide with tasks such as natural language understanding, content generation, programming assistance, and multimodal reasoning. From personal productivity tools to enterprise AI services, LLMs are reshaping the way people interact with information and technology.

Early advances in LLMs were largely driven by scaling dense Transformer-based~\cite{vaswani2023attentionneed} models and continuously pushing the limits of model parameters. For example, GPT-1 had 117M parameters~\cite{radford2018improving}, followed by GPT-2 with 124M to 1.5B parameters~\cite{gpt2}, and GPT-3~\cite{brown2020language} with 175B parameters. These large-scale dense models achieved outstanding performance across a wide range of tasks by continuously increasing parameter counts. However, as the size of dense models grows, they place increasing demands on computational resources, inference latency, and deployment cost. Therefore, efficiently serving such models in real-world applications has become a critical problem that warrants immediate attention.

Given these growing concerns, increasing attention has been drawn to sparse model architectures,
particularly Mixture of Experts (MoE) models~\cite{mix-of-experts}. The release of DeepSeek R1~\cite{deepseekai2025deepseekr1incentivizingreasoningcapability} demonstrated that sparse MoE models can perform as intelligently as traditional dense models while using significantly fewer computational resources. In MoE models, only a small subset of expert networks (i.e., specialized subnetworks) are activated per token during inference, enabling the model to scale to trillions of parameters while keeping the per-token computation manageable. 
Following DeepSeek R1, models such as Mixtral~\cite{jiang2024mixtralexperts} and MoE variants of LLaMA-4~\cite{meta2024llama4} have also adopted MoE architectures, demonstrating the effectiveness of this approach in achieving strong performance and improved inference efficiency.

Although MoE architectures hold great promise, they also introduce new challenges in system design and serving. In particular, MoE models typically rely on expert parallelism, distributing experts across multiple GPUs or nodes. To further increase inference throughput and support large-batch processing, expert parallelism is often combined with data parallelism. While this layered parallelism enables efficient hardware utilization, current LLM serving frameworks still mainly adopt scheduling strategies originally designed for dense models. At the engine level, simple Round-Robin dispatching ignores load conditions such as prefix token volume and KV Cache usage, often leading to load imbalance and underutilization. At the request level, First-Come-First-Serve remains the default, which can cause head-of-line blocking and long-tail latency under heavy workloads. More critically, at the expert level, existing systems fail to account for inter-layer expert affinity, dependency patterns, and GPU-local hotspots, resulting in uneven workload distribution across experts and devices.

To address these challenges, we propose \textbf{Gimbal}, a multi-layer scheduling framework designed to stabilize and optimize MoE-based LLM serving, with improvements in the following three areas:

\begin{itemize}
    \item \textbf{Engine-Level Scheduling}: adopting strategies such as Shortest-Job-First and priority-aware aging to reduce long-tail latency and improve throughput.
    
    \item \textbf{Request-Level Scheduling}: employing load-aware and affinity-preserving dispatching that considers prefix tokens, KV Cache usage, and user stickiness, to reduce Time To First Token and Time Per Output Token.

    \item \textbf{Expert-Level Scheduling}: mitigating expert hotspots and strategically placing dependent experts to improve efficiency and balance workload across GPUs.
\end{itemize}
Although our evaluation focuses on a representative MoE model, Gimbal is not tied to any model-specific internals. It relies only on general routing and scheduling signals exposed by MoE inference frameworks, and is therefore applicable to a broad range of MoE architectures, including Mixtral, Switch Transformer, and DeepSeek-style models.

Overall, the proposed multi-layer scheduling framework delivers substantial performance improvements for MoE-based LLM serving. By jointly optimizing request-level prioritization, engine-level load-aware dispatching, and expert-level dependency-aware placement, our approach effectively reduces both queuing delay and execution inefficiency under diverse workloads. Experimental results demonstrate that Gimbal achieves up to a 17.8\% reduction in Time To First Token and a 13.3\% reduction in Time-Per-Output-Token compared to the state-of-the-art baseline, while maintaining comparable throughput. 

The remainder of the paper is organized as follows: Section~\ref{sec:background} introduces the fundamental concepts of LLM and MoE architectures and discusses existing serving frameworks and their limitations. Section~\ref{sec:sys} presents the design of Gimbal and elaborates on its multi-layer scheduling mechanisms. Section~\ref{sec:imp} describes the implementation of the Gimbal framework. Section~\ref{sec:eval} evaluates Gimbal’s performance under various workloads and compares it against baselines. In the next section, we review related work, and Section~\ref{sec:conclusion} concludes the paper and discusses directions for future research.

\section{Background and Motivation}\label{sec:background}

This section provides an overview of the foundations of large language models and sparse Mixture-of-Experts architectures, followed by a discussion of modern LLM serving frameworks and their scheduling mechanisms. We then outline the limitations of existing systems and highlight the key motivations that drive the design of Gimbal.

\subsection{LLM and MoE model Basics}

Modern large language models are built upon the Transformer architecture~\cite{vaswani2023attentionneed}, where each layer combines attention, feedforward networks, residual connections, and normalization. In self-attention, the hidden states are projected into queries ($Q$), keys ($K$), and values ($V$), and computes attention via scaled dot-product. During inference, LLMs follow an autoregressive decoding process with two phases: \emph{prefill}, which attends over the entire prompt, and \emph{decode}, which generates tokens step by step. To avoid recomputing history, modern systems cache the key/value tensors (KV Cache), greatly reducing per-step computation.

As model sizes continue to grow, scaling dense architectures becomes increasingly resource-intensive. To address this, Mixture-of-Experts architectures~\cite{mix-of-experts} were introduced, replacing dense feedforward layers with sparsely activated \emph{experts}, as stated earlier. A lightweight router selects only a small subset of experts (e.g., top-$k$ out of hundreds) for each token, allowing models to scale to a massive number of parameters while keeping per-token computation manageable. 

\subsection{LLM Serving Architecture and Scheduling}
\begin{figure}[t]
\centering
    \includegraphics[width=0.9\columnwidth]{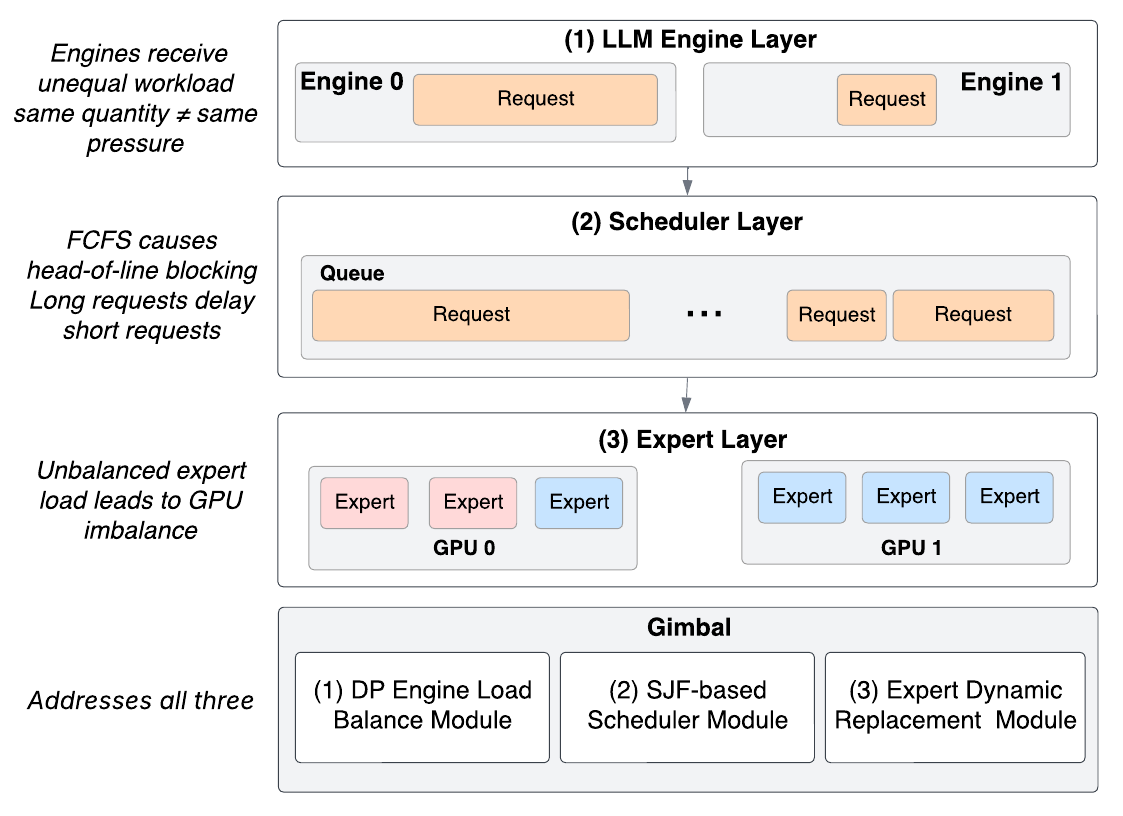}
    \caption{Current frameworks and their issues}  
    \label{fig: Current framework and issues}
\end{figure}

To meet the demands of real-time, high-throughput inference, modern LLM serving systems such as vLLM~\cite{vllm}, SGLang~\cite{zheng2024sglang}, TGI~\cite{tgi}, and Preble~\cite{srivatsa2024preble} typically adopt a layered architecture~(Figure~\ref{fig: Current framework and issues}) consisting of two major components: a global router and one or more inference engines. The router is responsible for distributing incoming requests across available engine replicas, while each engine performs token-level inference, including model execution, key-value (KV) cache management, and batching.

Within this architecture, scheduling can be viewed at multiple levels. At the \textit{engine level}, scheduling determines how requests are dispatched across different engine replicas in data-parallel (DP) serving. At the \textit{request level}, the engine manages concurrent requests internally, organizing them into dynamic batches and coordinating prompt processing and token decoding. With the introduction of expert parallelism (EP) in MoE models, an additional dimension of scheduling arises: \textit{expert} selection and routing. Here, tokens within a batch may activate different experts, and the system must determine how tokens are distributed to expert replicas deployed across GPUs or nodes.
This layered perspective has enabled modern LLM serving systems to improve hardware utilization and inference scalability, but it also introduces new challenges in coordinating scheduling across different levels.

\subsection{Motivation}
Currently, mainstream LLM serving systems such as vLLM~\cite{vllm} and SGLang~\cite{zheng2024sglang} mainly adopt service frameworks similar to Figure~\ref{fig: Current framework and issues} designed for traditional dense models, lacking specific optimizations for the characteristics of MoE models. In dense model scenarios, scheduling is relatively simple, often performed on a single machine or single GPU using FCFS, or via simple Round-Robin (RR) strategies across multiple inference \emph{engines} in multi-GPU settings. While these approaches are simple and easy to implement, they are insufficient for MoE serving scenarios: First, \emph{engine}-level RR scheduling is unaware of each \emph{engine}'s current load status, such as KV Cache usage, which can result in some \textit{engines} becoming overloaded while others remain underutilized, leading to suboptimal throughput and latency. 

In addition, they overlook opportunities to optimize prefix-cache reuse or leverage user stickiness by routing requests from the same user to the same engine. Queries from a single user are often logical, sequential, and contextually related, making them more likely to share common prefixes. Therefore, assigning engines based on historical request patterns or session affinity can significantly improve prefix-cache hit rates.
Second, request scheduling within each engine still commonly follows FCFS. While fair, FCFS can cause long-tail latency when a long request blocks subsequent short requests, degrading user experience. Under high-concurrency scenarios, FCFS also fails to fully utilize computational resources, resulting in low efficiency.
\begin{figure}[t] 
\centering
    \includegraphics[width=0.9\columnwidth]{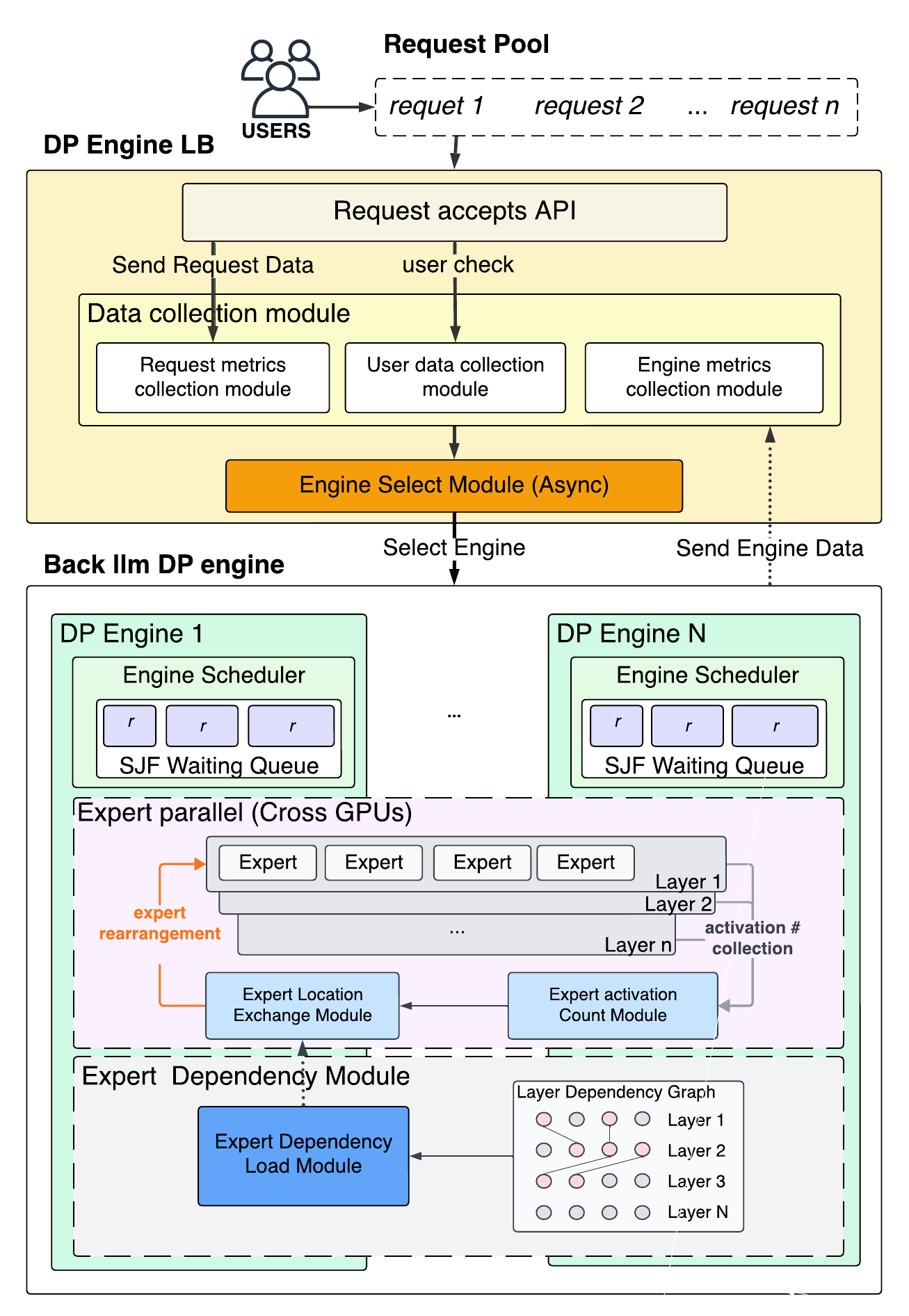}
    \caption{Gimbal's System Architecture}  
    \label{fig: System Architecture}
\end{figure}
More importantly, current inference frameworks often fail to adequately address inter-layer expert affinity and dependency, as well as localized hotspot issues on GPUs, leading to imbalanced expert workloads across GPU devices.
To address these challenges, we propose \textbf{Gimbal}, a multi-layer scheduling framework tailored for MoE-based LLM serving.

\section{System Design}\label{sec:sys}

This section provides an overview of the overall system architecture of Gimbal, as well as the specific functions, algorithms, and modeling techniques used in each component.

\subsection{System Architecture}
The system architecture, as shown in Figure~\ref{fig: System Architecture}, consists of three main components: 1) the DP Engine Load Balance, 2) SJF Scheduler, and 3) Expert Dynamic Replacement Module.
Gimbal begins with a global request-layer Data Parallel (DP) engine load balancer that receives incoming requests from the user request pool. The balancer performs lightweight data collection, including request features, user-level information, and real-time engine metrics, and asynchronously forwards these data to the engine selection module. Based on the current backend engine states and collected metrics, the balancer assigns each request to the most suitable DP engine.

Within each DP engine, Gimbal employs a Shortest-Job-First based scheduler. Each DP engine maintains its own SJF waiting queue, where requests are prioritized according to their token count. This design reduces tail latency, improves overall system responsiveness, lowers Time To First Token, and minimizes average queuing delay.

The final component of Gimbal is an Expert Dynamic Replacement Module that integrates both cross-layer expert dependencies and expert hot-spot exchanges. This module dynamically adjusts expert placement across GPUs during inference, preserving cross-layer expert affinity, balancing expert activation load, and minimizing inter-GPU communication whenever possible.

\subsection{DP Engine Load Balancing module}
\begin{algorithm}[t]
    \caption{DP Engine Selection Algorithm}
    \label{alg: DP Engine Selection Algorithm}
    \textbf{Input:} \( request \), \( dp\_engines[] \), 
    per-engine metrics: \( KVusage[n] \), \( RunningLoad[n] \), 
    \( user\_engine\_map \), thresholds \( \theta_{kv}, \theta_{dif\!f}, \theta_{load} \)

    \textbf{Output:} Selected engine \( e^* \)
    \begin{algorithmic}[1]
        \State \textbf{Initialize} \( e^* \gets \text{next engine from } dp\_engines[] \)
        \If{Metric data available}
            \State $i_{max} \gets \arg\max_i KVusage[i]$
            \State $i_{min} \gets \arg\min_i KVusage[i]$
            \If{\( KVusage[i_{max}] \ge \theta_{kv} \)} 
                \If{\( KVusage[i_{max}] - KVusage[i_{min}] \ge \theta_{dif\!f} \)}
                    \State \( e^* \gets dp\_engines[i_{min}] \) 
                \Else
                    \State \( load_{max}, load_{min} \) from \( RunningLoad[] \)
                    \If{\( load_{max} - load_{min} > \theta_{load} \)} 
                        \State $i^* \gets \arg\min_i RunningLoad[i]$
                        \State $e^* \gets dp\_engines[i^*]$
                    \EndIf
                \EndIf
            \ElsIf{request has user identity}
                \If{user in \( user\_engine\_map \) and not expired}
                    \State \( e^* \gets user\_engine\_map[user] \)
                \EndIf
            \EndIf
        \EndIf
        \State \textbf{Update} \( user\_engine\_map[user] \gets (e^*,  user\_id ) \)
        \State \Return \( e^* \)
    \end{algorithmic}
\end{algorithm}
The DP engine load balancing (LB) module is responsible for assigning each incoming request to the most appropriate data-parallel (DP) engine based on real-time backend states and user-level characteristics. Algorithm~\ref{alg: DP Engine Selection Algorithm} outlines the decision-making workflow.

The algorithm begins by directly selecting an initial candidate engine from the ordered list \( dp\_engines[] \), ensuring that the process can proceed even in the absence of metric data (Algorithm~\ref{alg: DP Engine Selection Algorithm} line~1).
When inference begins, each LLM engine automatically computes both its total available KV-cache capacity and the amount of KV cache currently consumed, reported as \( KVusage[n] \). Meanwhile, the scheduler within each backend engine tracks its ongoing computational load, including the number of running and waiting tokens, which we collectively denote as \( RunningLoad[n] \). These two sets of metrics are asynchronously delivered to the Engine Load Balancer (LB) module through a message queue. Upon receiving them, the load balancer first evaluates the key--value (KV) cache usage of each engine (Algorithm~\ref{alg: DP Engine Selection Algorithm} lines~2 to 5).
If the maximum KV usage exceeds the saturation threshold~$\theta_{kv}$, the module checks whether the imbalance between engines is significant.

$\theta_{dif\!f}$ denotes the difference between the maximum and minimum KV-cache usage percentages among all engines, capturing the degree of cross-engine imbalance. Since the actual KV cache utilization rates of each engine are not absolutely equal in actual inference, this threshold needs to be set to allow for a certain tolerance before judging the system as unbalanced.
When the difference exceeds~$\theta_{dif\!f}$, the request is routed to the engine with the lowest KV usage, reducing the risk of out-of-memory issues and distributing prefill load more evenly (Algorithm~\ref{alg: DP Engine Selection Algorithm} lines~6 to 7).

If no severe KV imbalance is detected, the scheduler considers the runtime load \( RunningLoad[n] \) of each engine (Algorithm~\ref{alg: DP Engine Selection Algorithm} lines~8 to 14). By using the request's token count as the workload metric, the scheduler alleviates the imbalance that would arise from measuring load solely by the number of requests. 
If the difference between the \( RunningLoad[n] \) and the engine's running load is greater than $\theta_{load}$, the system considers the load between the engines to be mismatched, and the request will be directed to the engine with the lowest running load, thereby improving throughput and reducing queuing latency (Algorithm~\ref{alg: DP Engine Selection Algorithm} lines~9 to 12).
It is recommended to set $\theta_{load}$ according to the distribution of the input token count of the specific inference request to allow for small load differences comparable to a single typical request.

When the incoming request contains a user identity, the module further attempts to enable optional ``user affinity'' (Algorithm~\ref{alg: DP Engine Selection Algorithm} lines~15 to 17).
Since pure user stickiness may cause requests to be repeatedly assigned to the same engine due to cached user–engine mappings—potentially accelerating KV-cache imbalance—the affinity mechanism is only applied when no engine shows KV overuse, and the user's most recent mapping exists and has not expired. As demonstrated by SGLang~\cite{zheng2024sglang}, prefix caching can significantly reduce GPU computation and memory overhead. Because consecutive requests from the same user are more likely to reuse previously cached prefixes, we attempt to maximize prefix-cache reuse and reduce GPU computation by incorporating user affinity into scheduling.

After the final engine assignment is computed, the module updates the user-to-engine mapping using the current user identifier $ user\_id$  and the selected engine, and then returns \( e^* \) (Algorithm~\ref{alg: DP Engine Selection Algorithm} lines~21 to 22).

\subsection{SJF Scheduler module}

The SJF Scheduler module runs inside each DP engine and is responsible for reordering pending requests before each forward pass. Algorithm~\ref{alg:SJF with Aging} summarizes the scheduling workflow. 

\begin{algorithm}
    \caption{Request Scheduler with SJF and aging}
    \label{alg:SJF with Aging}
    \textbf{Input:} \( waiting\_queue\), \( time_{now} \), threshold \( \theta_{age} \)

    \textbf{Output:} re-order \( waiting\_queue' \)
    \begin{algorithmic}[1]
        \For{each request \( r \in waiting\_queue \)}
            \State waiting time \( w_r = time_{now} - r.arrival\_time \)
            \If{\( w_r \ge \theta_{age} \)}
                \State Assign \textbf{high priority} to \( r \) 
            \Else
                \State Assign priority based on request's prefill length \( r.prompt \)
            \EndIf
        \EndFor
        \State Sort \( waiting\_queue\) by priority ascending
        \State \Return \( waiting\_queue' \)
    \end{algorithmic}
\end{algorithm}

Within the scheduler’s \( waiting\_queue \), we implement an approximate shortest-job-first policy (Algorithm~\ref{alg:SJF with Aging} lines~1 to 7). We use the prefill token count \( r.prompt \) of each request as the priority metric (shorter first). This choice avoids the unreliable output-length prediction used in related work~\cite{fu2024efficient,qiu2024efficientinteractivellmserving,zhao2025seallmserviceawarelatencyoptimizedresource,wu2024fastdistributedinferenceserving}, and leverages the compute-intensive nature of the prefill phase, providing a stable and model-agnostic estimate of request cost.

To prevent starvation of large requests, the scheduler employs an aging mechanism: if the waiting time of a request exceeds the threshold~$\theta_{age}$, it is promoted to high priority regardless of size. Finally, the waiting queue is sorted according to the assigned priorities and returned for execution (Algorithm~\ref{alg:SJF with Aging} lines~9 to 10). The recommended value for $\theta_{age}$ is a value lower than the P99 TTFT during normal inference.

\subsection{Expert Dynamic Replacement Module Design}


MoE models often suffer from load imbalance in expert-parallel execution, as expert routing often concentrates computation on a small subset of experts while leaving others underutilized. To better understand this behavior, we analyze the expert activation patterns of the Qwen3-30B-A3B model~\cite{qwen3} during inference.

\begin{figure}[h] 
\centering
    \includegraphics[width=\columnwidth]{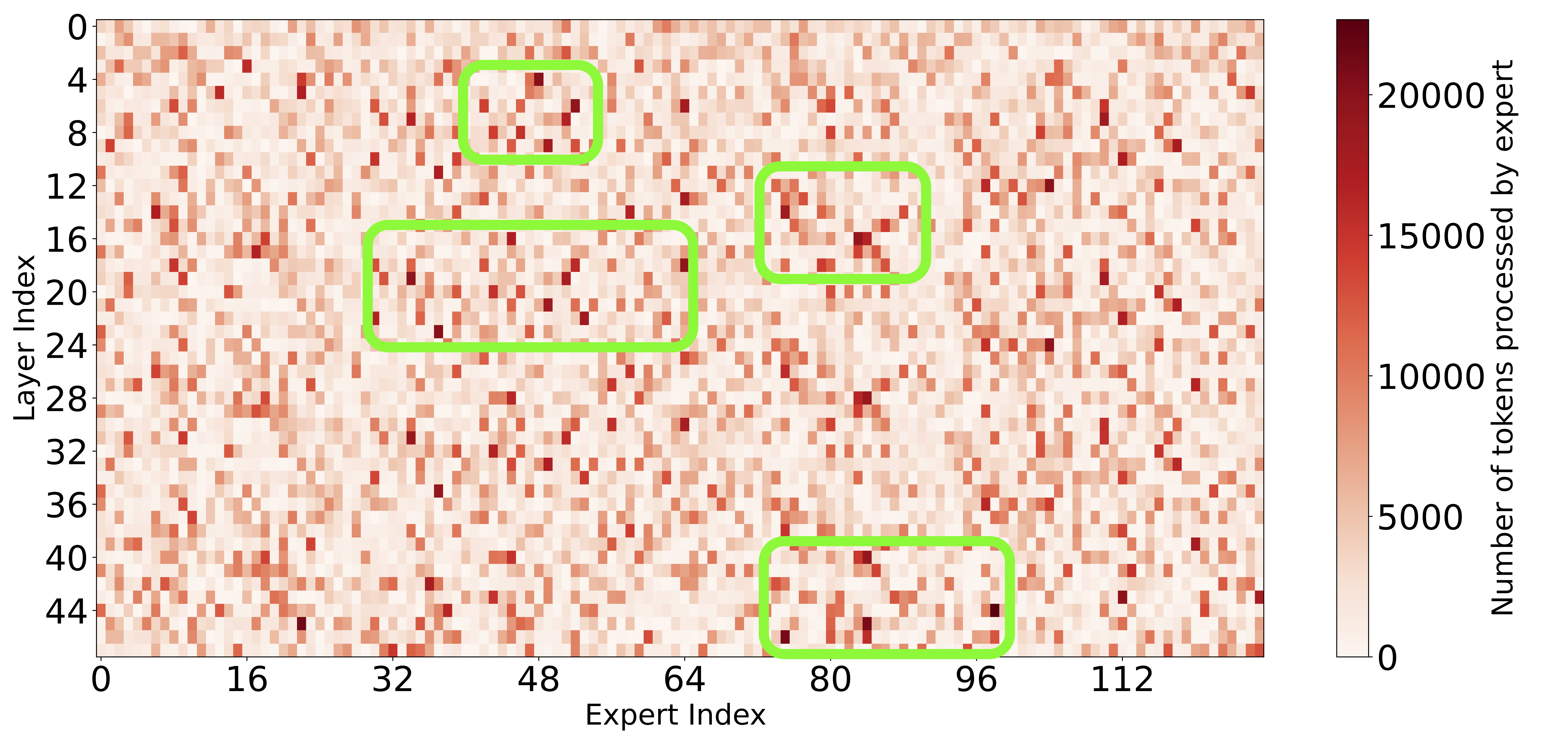}
    \caption{Expert Heat Map (Expert activation is not balanced)}  
    \label{fig: Expert Heat Map}
\end{figure}

Figure~\ref{fig: Expert Heat Map} illustrates the expert activation distribution of the model over 200 inference requests sampled from vLLM's random dataset. Using a random dataset helps avoid bias introduced by specific prompts, contextual patterns, or symbol repetitions, ensuring that the observed expert activation behavior reflects the intrinsic routing characteristics of the model.
We observe that several layers—such as Layer 16, 18, 44, 45, and 46—exhibit noticeable expert imbalance, where a small subset of experts is activated disproportionately often while many others remain largely inactive. Such skewed activation severely exacerbates GPU imbalance in expert-parallel execution, since uneven expert load directly translates into imbalance cross-GPU computation.

To address this issue, Guo et al.~\cite{deepseekai2025deepseekr1incentivizingreasoningcapability} introduced the Expert Parallel Load Balancing (EPLB) mechanism, which identifies expert hotspots by tracking expert activation frequency and then redistributes experts across GPUs to mitigate load imbalance. However, conventional EPLB relies solely on activation counts when determining placement. Recent studies, including MoeTuner~\cite{go2025moetuner} and Inter-Layer Expert Affinity~\cite{yao2024exploiting}, reveal that MoE models exhibit strong inter-layer expert affinity, i.e, certain experts in one layer consistently activate specific experts in the next. Exploiting these cross-layer dependencies phenomena, we design a more advanced replacement algorithm that yields substantial performance benefits.

\begin{figure}[h] 
\centering
    \includegraphics[width=\columnwidth]{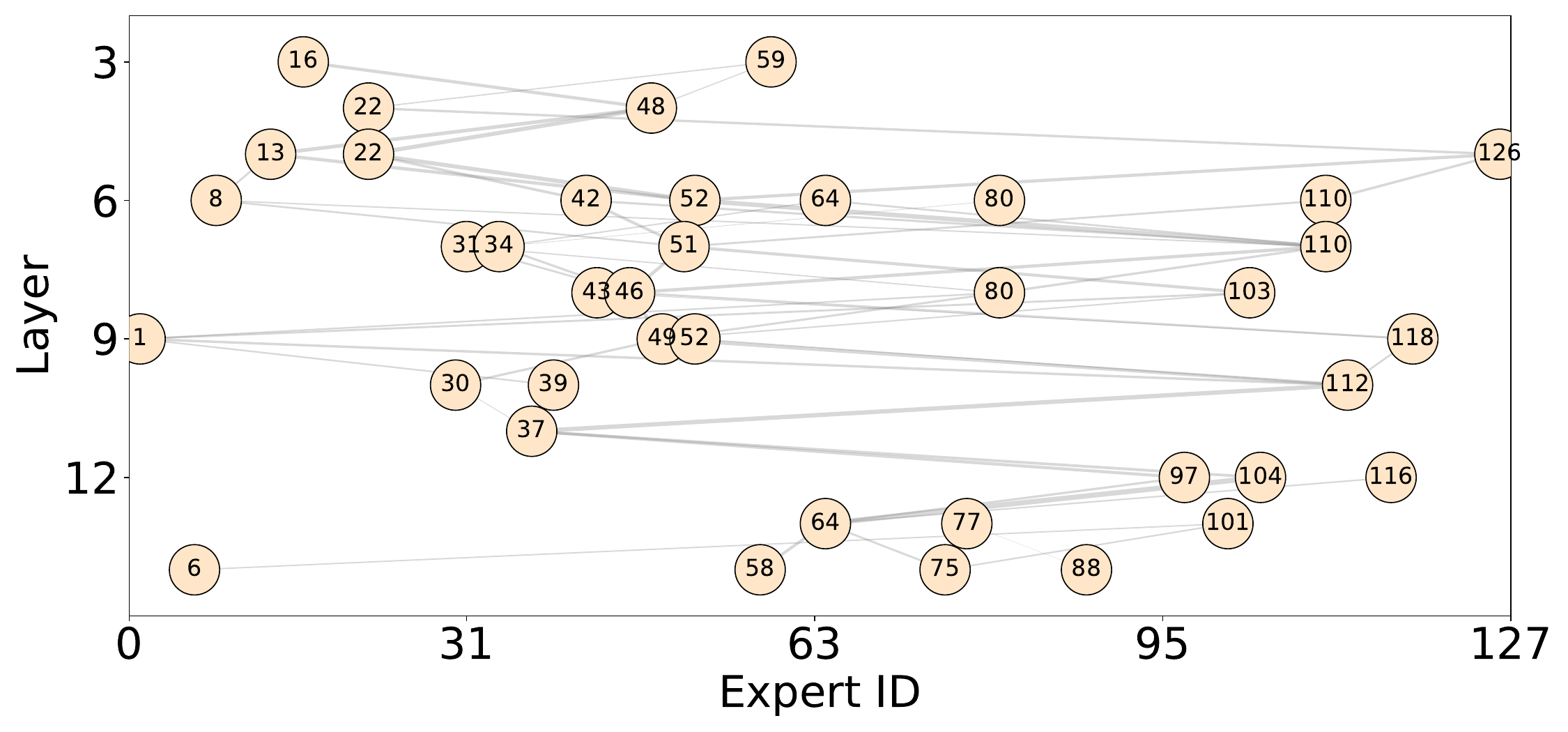}
    \caption{Partial layer's Expert affinity}  
    \label{fig: expert affinity}
\end{figure}

In collecting the earlier expert activation heatmap, we also recorded how each token in the 200 requests traversed all MoE layers and which experts it selected. As shown in Figure~\ref{fig: expert affinity}, we retain and visualize only expert-to-expert connections that occurred more than 100,000 times. From this, we observe that the Qwen3-30B-A3B model also shows strong cross layer expert affinity, where experts in one layer strongly tend to activate only a small subset of experts in subsequent layers.

Motivated by these observations, we integrate both optimization directions and introduce our Expert Dynamic Replacement Module. Instead of relying solely on activation counts, our method incorporates inter-layer affinity during dynamic expert relocation. By co-locating affinity-linked experts on the same GPU, the system reduces cross-GPU communication overhead and lowers per-token inference latency, while simultaneously alleviating expert hotspot imbalance across GPUs.

\subsubsection{Problem Formulation}

We formulate the expert placement problem as a multi-constraint graph partitioning task. 
Consider an MoE model with $m$ experts and $n$ transformer layers. Let 
\begin{align}
A \in \mathbb{R}_{\ge 0}^{n \times m}
\end{align}
be the activation matrix, where $A_{i,j}$ denotes the activation intensity of expert $j$ at layer $i$. 
Let $g<m$ be the number of available GPUs, and assume each GPU hosts approximately $m/g$ experts.

Communication between consecutive layers is characterized by nonnegative weights $E_{i,j,k}$, representing the volume of traffic from expert $j$ in layer $i$ to expert $k$ in layer $i{+}1$. For convenience, we define the aggregated inter-expert communication weight:
\begin{align}
W_{j,k}=\sum_{i=1}^{n-1} E_{i,j,k},
\end{align}
which captures the total cross-layer dependency between experts $j$ and $k$.

The goal is to partition the $m$ experts into $g$ groups (GPUs) such that:
(i) for every layer $i$, the activation load assigned to each GPU is balanced, and 
(ii) the total cross-GPU communication cost induced by $W_{j,k}$ is minimized.

\subsubsection{MILP Formulation}

We introduce binary variables $x_{j,p}\in\{0,1\}$, where $x_{j,p}=1$ indicates that expert $j$ is assigned to GPU (partition) $p\in\{1,\dots,g\}$.

\paragraph{Assignment and size constraints.}
Each expert must be assigned to exactly one GPU:
\begin{align}
\sum_{p=1}^g x_{j,p} = 1 \qquad \forall j=1,\dots,m.
\end{align}
Each GPU hosts exactly $m/g$ experts:
\begin{align}
\sum_{j=1}^m x_{j,p} = m/g \qquad \forall p=1,\dots,g.
\end{align}

\paragraph{Row-wise balance}
For layer $i$, define the total activation load:
\begin{align}
T_i = \sum_{j=1}^m A_{i,j},
\end{align}
and the \emph{ideal balanced load} per GPU:
\begin{align}
L_i = \frac{T_i}{g}.
\end{align}
The actual load assigned to GPU $p$ at layer $i$ is:
\begin{align}
L_{i,p} = \sum_{j=1}^m A_{i,j} x_{j,p}.
\end{align}
We impose a maximum deviation $D\ge 0$ between actual and ideal load:
\begin{align}
L_{i,p}-L_i &\le D, \qquad \forall i,p, \\
L_i-L_{i,p} &\le D, \qquad \forall i,p.
\end{align}
The parameter $D$ specifies the maximum deviation allowed between the ideal balanced activation load $L_i$ and the actual load $L_{i,p}$ assigned to each GPU at layer $i$. A smaller value of $D$ enforces strict load balance, ensuring that every GPU receives nearly equal activation volume at each layer. However, such strictness may require excessive expert relocation and can interfere with desirable co-location of affinity-linked experts. Conversely, a larger $D$ relaxes the balance constraint, allowing certain GPUs to host heavier activation loads in exchange for improved communication locality.

\paragraph{Communication cut.}
Define same-part indicator $s_{j,k,p}\in\{0,1\}$ to denote whether experts $j$ and $k$ are both placed on GPU $p$. This is linearized using:
\begin{align}
s_{j,k,p} \le x_{j,p}, \qquad 
s_{j,k,p} \le x_{k,p}, \qquad 
s_{j,k,p} \ge x_{j,p} + x_{k,p} - 1.
\end{align}
An expert pair contributes to the cut if they are placed on different GPUs. The total communication cut cost is:
\begin{align}
\mathrm{Cut} 
&= \sum_{j<k} W_{j,k}
\left(1 - \sum_{p=1}^g s_{j,k,p}\right).
\end{align}

\paragraph{Objective.}
We minimize a weighted combination of row-wise imbalance and communication cut:
\begin{align}
\min \; 
\alpha D 
\;+\;
\beta \sum_{j<k} W_{j,k}
\left(1 - \sum_{p=1}^g s_{j,k,p}\right),
\label{align:ojb}
\end{align}
where $\alpha,\beta>0$ control the tradeoff between activation-load balancing and communication minimization.

\subsubsection{Implementation}
Since the MILP formulation described above is computationally expensive and unsuitable for real-time inference, introducing huge overhead, we design an intuitive  heuristic expert replacement strategy (Algorithm~\ref{alg:affinity-eplb}) that efficiently balances expert load while minimizing communication overhead.

Our heuristic prioritizes the communication term by first co-locating interdependent experts according to the affinity matrix $\mathcal{M}$ (similar to the expert dependency patterns illustrated in Figure~\ref{fig: expert affinity}). Experts with strong inter-layer dependencies recorded in $\mathcal{M}$ are co-located on a single ``anchor'' GPU to minimize cross-GPU communication (Algorithm~\ref{alg:affinity-eplb} line 3). 
We observe that strong inter-layer expert dependencies in MoE models are typically sparse and localized. Modern MoE training objectives explicitly encourage balanced routing and discourage large-scale dense expert coupling, resulting in only a small number of top-$E$ expert pairs exhibiting strong and persistent dependencies. Consequently, the affinity matrix $\mathcal{M}$ is constructed to capture only the most communication-critical dependency relationships. If the capacity of a single anchor GPU becomes a concern, the construction of $\mathcal{M}$ can be made more selective by tightening the statistical threshold or reducing the top-$E$ criterion, thereby ensuring that only the strongest and most impactful expert dependencies are co-located, and that all dependency-linked experts retained in $\mathcal{M}$ can be accommodated on the designated anchor GPU.

The remaining experts are then distributed across all $g$ GPUs according to their activation intensity $A_{i,j}$ (Similar to Figure~\ref{fig: Expert Heat Map}'s expert activation heat map) using a greedy least-loaded placement policy, achieving balanced per-layer activation load while preserving communication locality (Algorithm~\ref{alg:affinity-eplb} line 4).

During execution, the system periodically re-evaluates the expert placement every $\tau$ steps. 
During each relocation, experts with strong affinities recorded in $\mathcal{M}$ are always placed on the designated anchor GPU, avoiding repeated migrations of dependency-linked experts and thereby reducing expert migration overhead and cross-device transfer cost. The anchor GPU index $k$ remains fixed and is manually specified before system startup. The remaining experts are greedily rebalanced across all GPUs according to recent activation statistics, maintaining load balance while preserving the communication locality achieved by the fixed-anchor placement.

\begin{algorithm}
\caption{Expert Dynamic Replacement Algorithm}
\label{alg:affinity-eplb}
\textbf{Inputs:} affinity matrix $\mathcal{M}$,\; activation matrix $A$,\; GPU index $k$,\; number of GPUs $g$,\; step interval $\tau$.\\
\begin{algorithmic}[1]

\Procedure{Exp-relocation}{$k$}
  \State \textbf{Affinity placement:} place all experts appearing in $\mathcal{M}$ on GPU $k$.
  \State \textbf{Greedy balancing:} assign the remaining experts to GPUs $0..g{-}1$
         by descending $A_{i,j}$ with a least-loaded-GPU policy.
\EndProcedure

\State \Call{Exp-relocation}{$k$} Run once during system loading.

\For{iteration step $t=1,2,\dots$}
  \If{$t \bmod \tau = 0$} 
    \State \Call{Exp-relocation}{$k$}
  \EndIf
\EndFor
\end{algorithmic}
\end{algorithm}

\section{Implementation}\label{sec:imp}
We implement Gimbal on top of the SOTA inference framework vLLM~\cite{vllm} (version~0.9.1), extending its internal scheduling and load-balancing mechanisms and adding our own system logic.

At the DP layer, we modify both the request dispatching logic and the metric–collection pipeline to support our DP Engine LB module (Algorithm~\ref{alg: DP Engine Selection Algorithm}). Specifically, we extend the ZeroMQ communication payload so that the DP Engine Load Balancer can asynchronously receive real-time backend metrics from each engine (e.g., KV-cache usage and running token counts), enabling more adaptive and fine-grained load balancing.

Inside each engine, we replace the default FCFS scheduler with a prefill-length–aware SJF strategy (Algorithm~\ref{alg:SJF with Aging}), which reduces queuing delay and improves tail latency.

Furthermore, we port and integrate Expert Parallel Load Balancing (EPLB) into our system to support dynamic expert relocation during inference. Through the algorithm and strategy shown in Figure~\ref{alg:affinity-eplb}, we integrate GPU affinity awareness into this module to ensure that experts with strong inter-layer dependencies are preferentially located on the same GPU, thereby reducing inter-GPU communication overhead and improving latency.

\section{Performance Evaluation}\label{sec:eval}

\begin{figure*}[t]
    \centering
    \begin{subfigure}[t]{0.191\textwidth}
    \centering
    \includegraphics[width=\textwidth]{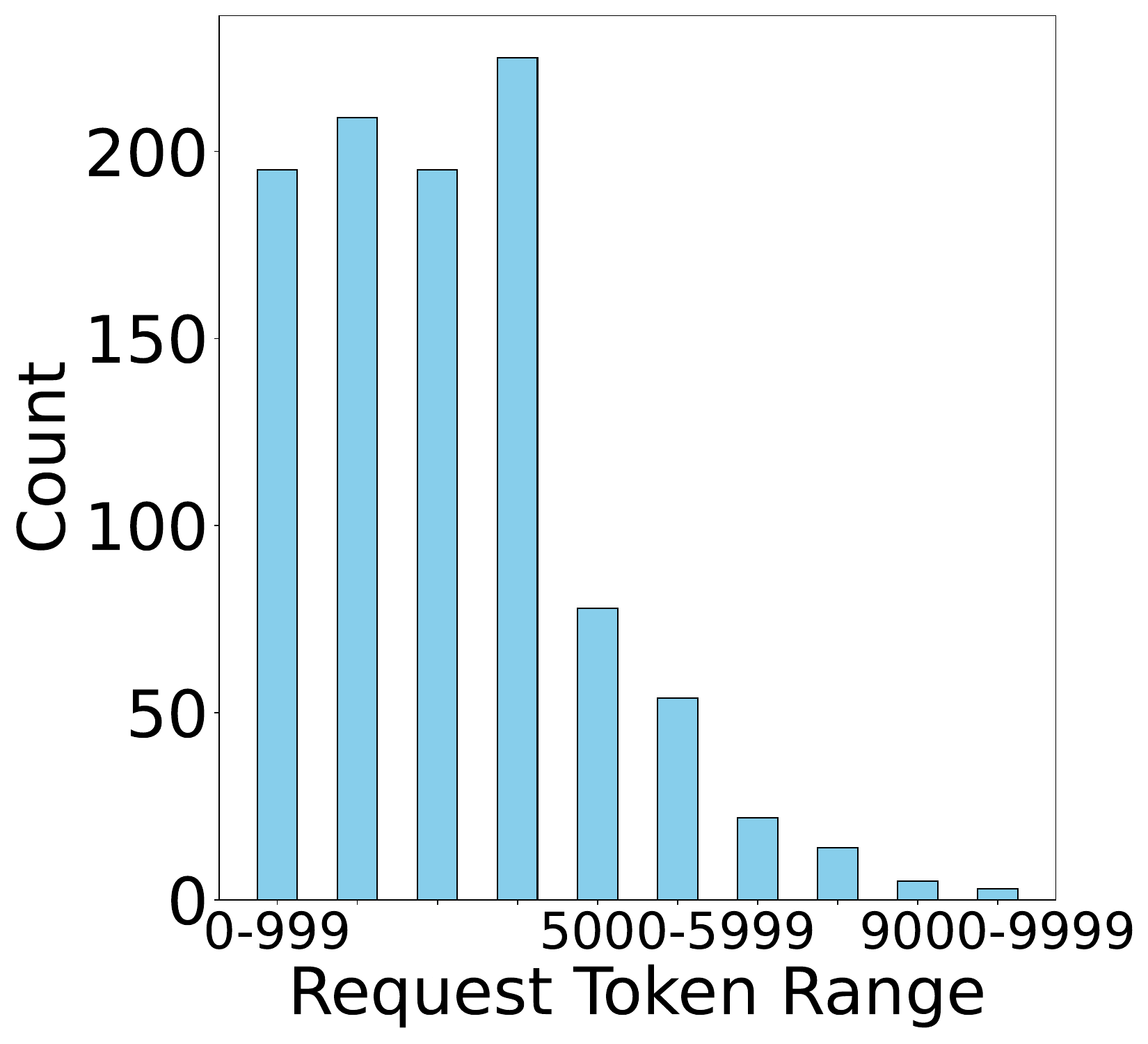} 
    \caption{Random}
    \label{random_distribution.pdf}
    \end{subfigure}
    \hfill
    \begin{subfigure}[t]{0.191\textwidth}
        \centering
        \includegraphics[width=\textwidth]{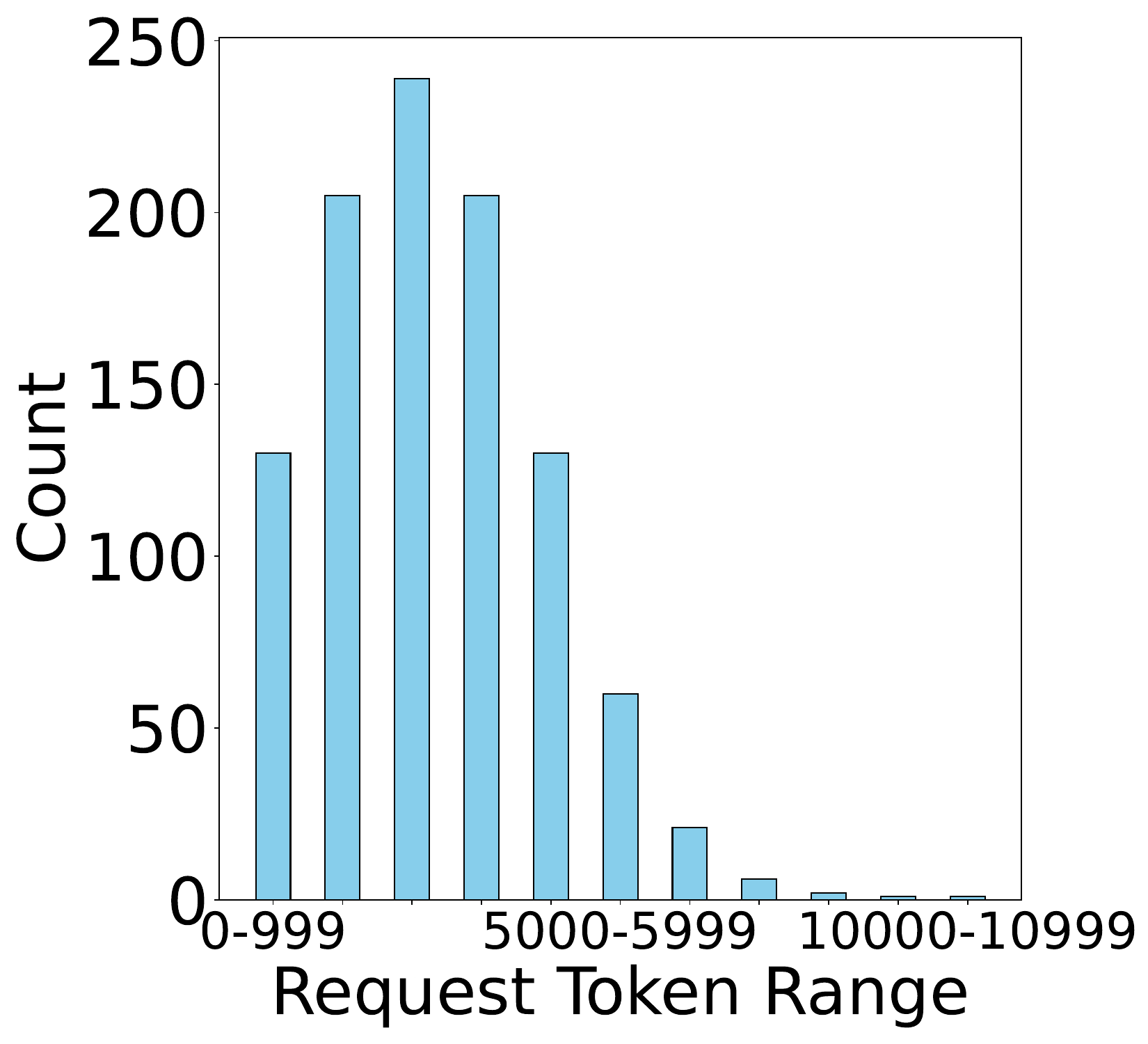}
        \caption{Central}
        \label{fig:Central_distribution.pdf}
    \end{subfigure}
    \hfill
    \begin{subfigure}[t]{0.191\textwidth}
        \centering
        \includegraphics[width=\textwidth]{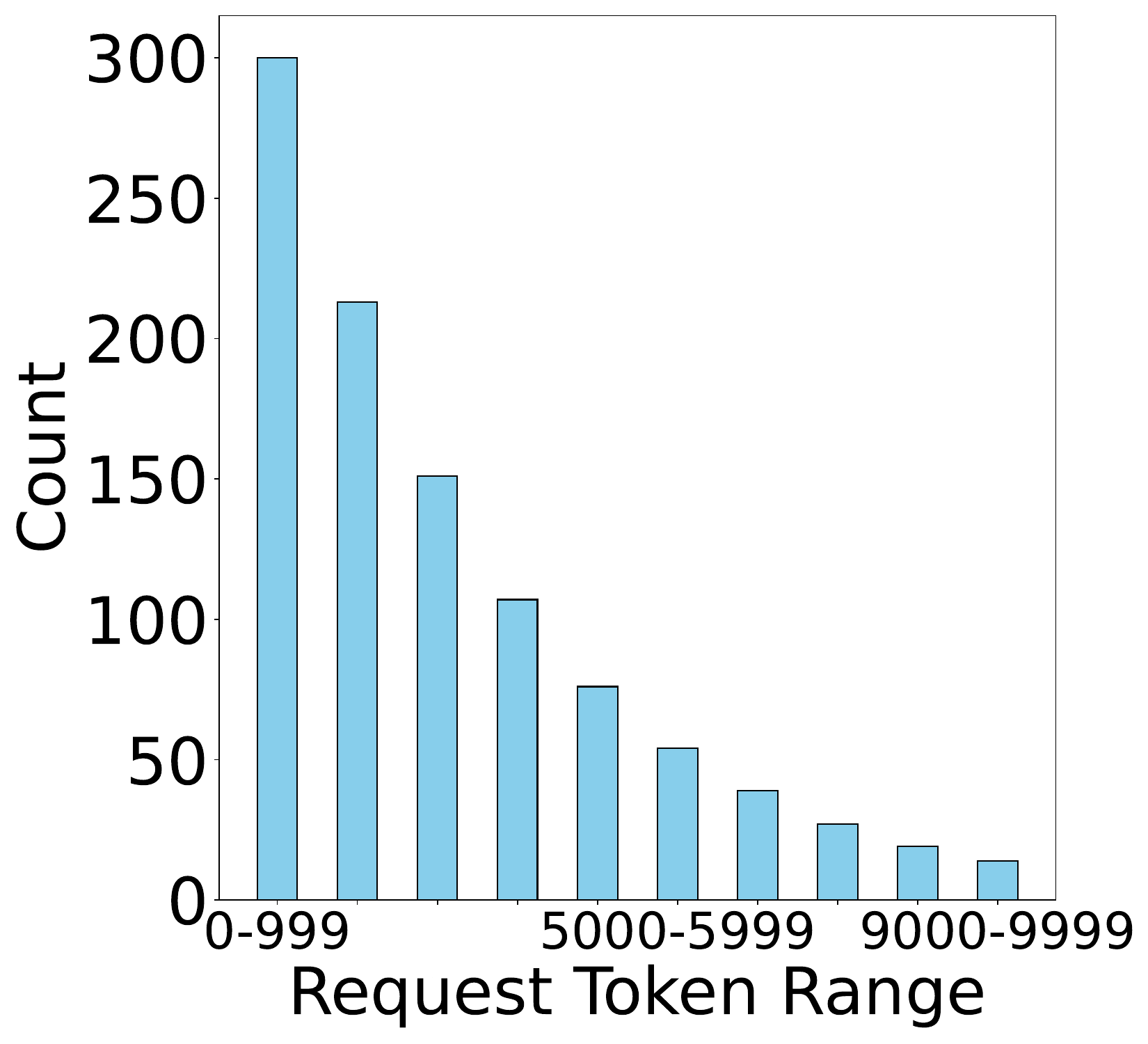}
        \caption{Descending}
        \label{fig:Descending_distribution.pdf}
    \end{subfigure}
    \hfill
    \begin{subfigure}[t]{0.191\textwidth}
        \centering
        \includegraphics[width=\textwidth]{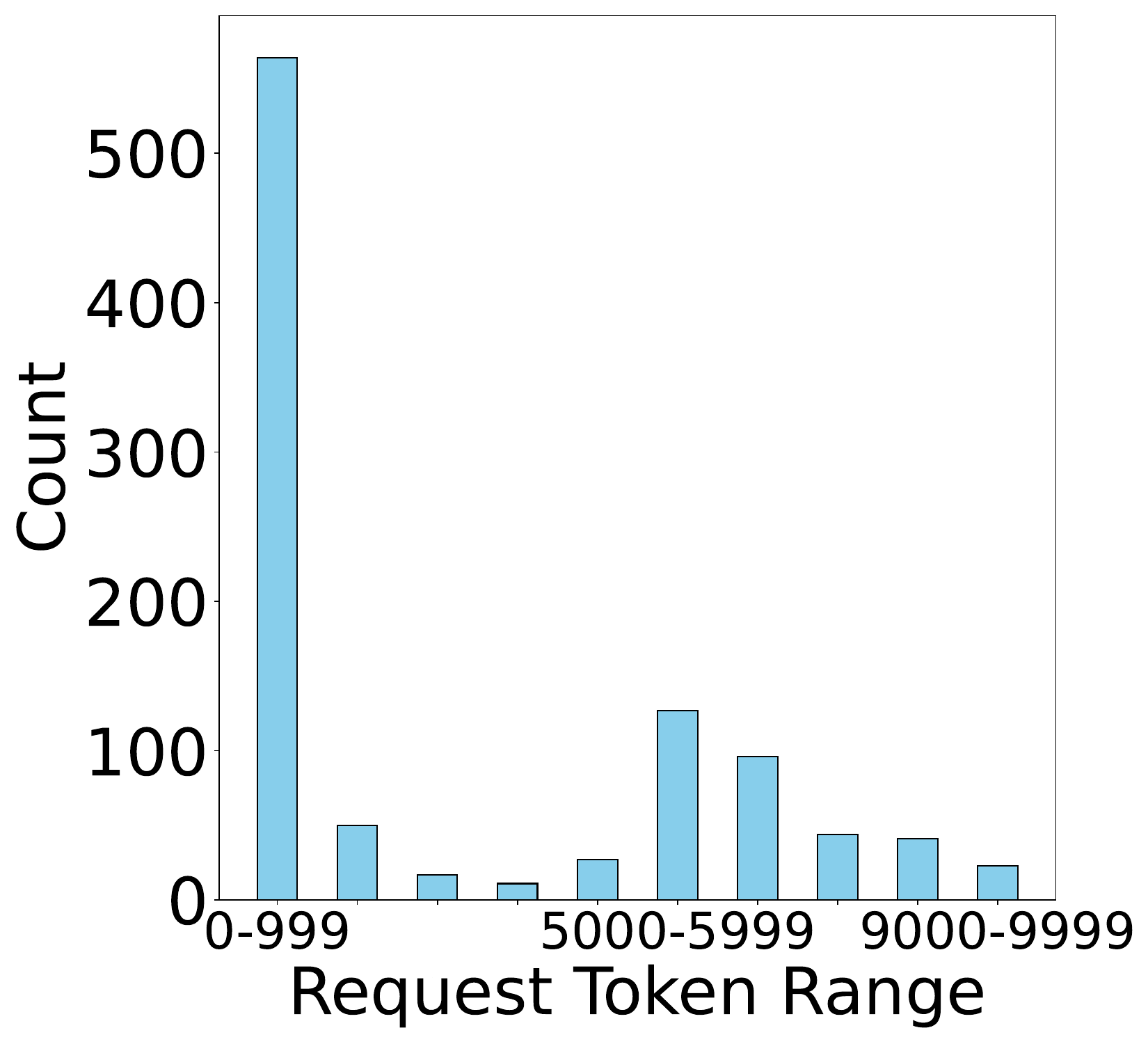}
        \caption{Two-end}
        \label{fig:Two-end_distribution.pdf}
    \end{subfigure}
    \hfill
    \begin{subfigure}[t]{0.191\textwidth}
        \centering
        \includegraphics[width=\textwidth]{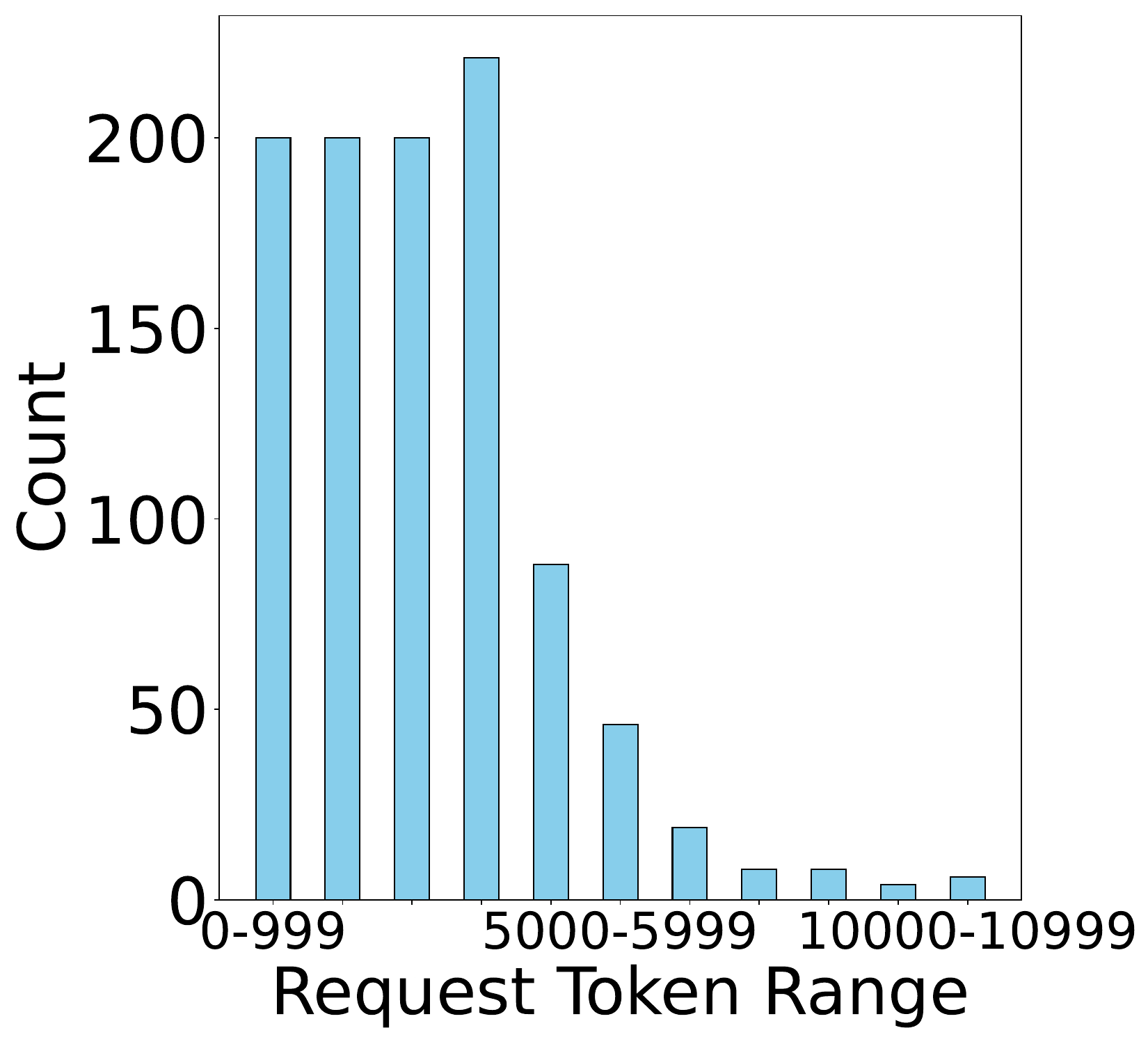}
        \caption{Average}
        \label{fig:Average_distribution.pdf}
    \end{subfigure}
    \caption{Burst dataset distribution}
    \label{fig:Burst dataset distribution}
\end{figure*}

\subsection{Experimental Setup}
\subsubsection{\textbf{Testbed}}
All experiments are conducted on the server equipped with two Intel (R) Xeon (R) Gold 6326 CPUs running at 2.90GHz, two NVIDIA A100 80GB GPUs interconnected via NVLink, and 1~TB of system memory. This hardware configuration provides high GPU memory capacity and fast inter-GPU bandwidth, allowing us to evaluate MoE inference under realistic multi-GPU environments. 
For other advanced features of vLLM, such as chunked prefill and continuous batching, we keep the default settings. We enable expert parallelism so that all experts in each MoE layer are distributed across all GPUs. In addition, we use pplx-kernels~\cite{pplx-kernels} as the communication backend for MoE all-to-all exchanges to ensure efficient point-to-point cross-GPU communication among experts.

\subsubsection{\textbf{Threshold settings  and Rationale}}

\begin{itemize}
    \item $\theta_{kv}$: most open-source frameworks use a maximum of 90\% of GPU memory by default, so we adopt a similar limit, setting $\theta_{kv}$ to 0.9. That is, if the KV cache usage of a certain engine exceeds 90\%, it is considered to have exceeded the threshold.
    \item $\theta_{dif\!f}$, in practice, KV-cache usage across engines rarely aligns perfectly, so we set $\theta_{dif\!f}$  to 10\%, allowing a reasonable tolerance before classifying the system as imbalanced.
    \item $\theta_{load}$: We use the BurstGPT dataset~\cite{burstgpt} for our experiments, in which 97.6\% of requests contain no more than 3000 tokens. Accordingly, we set $\theta_{load}$ to 3000. Requests exceeding this threshold typically indicate a significant imbalance in prefill workload, where the most-loaded engine and the least-loaded engine differ by at least one additional large request in their queues. In such cases, directing the incoming request to the engine with the lowest current load helps mitigate prefill skew, thereby improving overall throughput and reducing queuing latency.
    \item $\theta_{age}$: we set $\theta_{age}$ to 5 seconds. Our environment consistently has a P99 TTFT (99\% of the TTFT time is below 4900 milliseconds) when performing high-load (Request-rates of 1.4 RPS); therefore, any request with a wait time exceeding 5 seconds can be considered an anomalous latency and should be prioritized.
    \item $\tau$: currently, open-source frameworks that support EPLB, such as vLLM and SGLang, all perform expert re-migration by default every 3000 steps, so we will maintain this consistency here.
\end{itemize}

\subsubsection{\textbf{LLM Model}}
Considering the compute constraints of our testbed and our objective to study MoE-specific scheduling behavior, we select the Qwen3-30B-A3B model~\cite{qwen3}, one of the most representative and state-of-the-art small/medium-scale MoE models.
\subsubsection{\textbf{Traces}}
To evaluate performance under different load conditions, we use the BurstGPT dataset~\cite{burstgpt}, a popular open-source dataset, commonly used in the literature \cite{zhang2025blitzscale}, \cite{kim2025oaken}.
During performance experiments, we disable prefix caching to avoid interference and measurement bias. Following recent studies~\cite{flexpipe, aegaeon}, we adopt a similar methodology by modifying the distributional shape of the BurstGPT traces to evaluate robustness under workload variations and to minimize dataset-induced bias. Specifically, we partition the BurstGPT workload into five representative distributions—Random, Central, Descending, Two-end, and Average (Fig.~\ref{fig:Burst dataset distribution}). We sample 1,000 requests from the BurstGPT dataset according to each distribution, using random seeds as experiment's input.
Since BurstGPT does not contain user identifiers, we additionally use the ShareGPT dataset~\cite{sharegpt} to specifically evaluate user-affinity behavior and prefix-cache reuse efficiency.

\subsubsection{\textbf{Metrics}}
We evaluate several key metrics commonly used in LLM serving:
\begin{itemize}
    \item TTFT, measuring the latency from sending a request to receiving the first generated token (including prefill/prompt processing);
    \item TPOT, representing the average decoding latency per output token, excluding the first generated token.
    \item Prefix Cache Block Hit Count: the total number of KV-block matches across all DP engines during the entire experiment. 
    This metric reflects the absolute amount of KV-block reuse enabled by user-affinity scheduling.
    \item Prefix Cache Hit Rate (global): the ratio between the number of cache-hit KV blocks and the total number of probed KV blocks across all engines. 
    It reflects the overall probability that an accessed KV block can be reused from the global prefix-cache table.
\end{itemize}

\subsubsection{\textbf{EPLB Affinity Matrix}}
Our Expert Dynamic Replacement Module requires two types of input: 1) expert activation counts, and 2) the expert inter-layer dependency matrix. Expert activation counts are collected during inference using built-in vLLM's EPLB measurement logic, which records the total number of activations per expert at each level over a period of time.
For the expert inter-layer dependency matrix, we modified vLLM code to allow the system to record, for each activation of an expert in the upstream layer, which expert is selected in the downstream layer. Then, we used the built-in random input data generation function of vLLM inference framework to run benchmark tests to collect cross-layer expert affinity. Using a random data avoids dataset-specific bias and better reveals the true dependency relationships between experts.  The matrix includes all layers, representing the complete cross-layer dependency matrix covering all experts with dependency attributes from layer 0 to layer 47, similar to the sampled pattern in Figure~\ref{fig: expert affinity}. 

\subsubsection{\textbf{Baselines}}
We compare Gimbal with vLLM\footnote{vLLM(0.9.2) \url{https://github.com/vllm-project/vllm}.} as well as three ablated variants of our proposed framework. Each ablation isolates the effect of each of the proposed components: 
\begin{itemize}
\item  \textbf{DPLB}: Only the DP Engine Load Balancer is enabled.
\item  \textbf{SJFS}: Only the per-engine SJF scheduler is enabled.
\item  \textbf{EDR}: Only the Expert Dynamic Replacement Module is enabled.
\end{itemize}
This decomposition allows us to quantify the contribution of each module within the overall multi-layer scheduling design.

\subsection{Experimental Results}

\subsubsection{\textbf{TTFT Results}}

\begin{figure*}[tp]
\centering
    \begin{subfigure}[t]{0.191\textwidth}
    \centering
    \includegraphics[width=\textwidth]{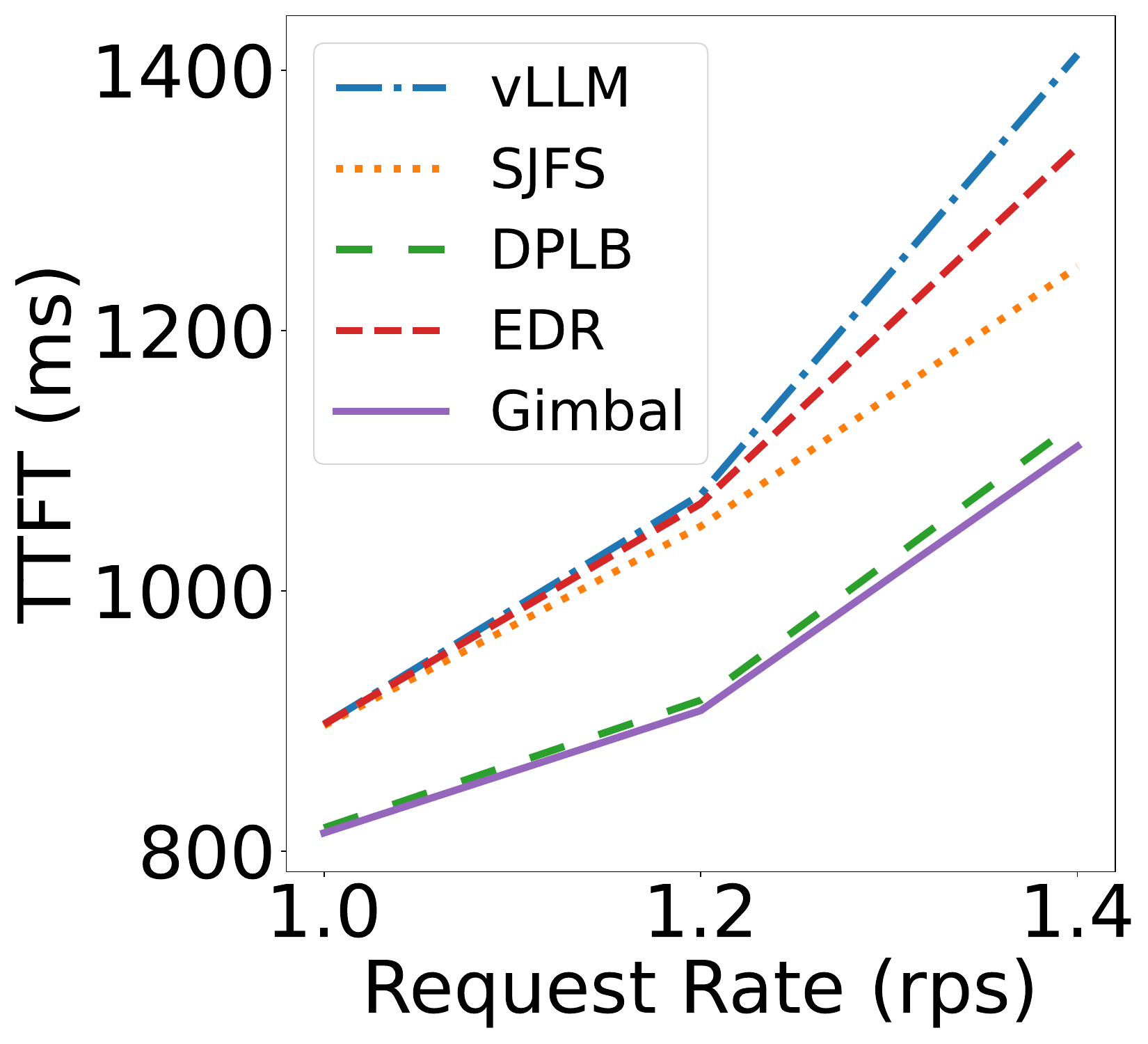} 
    \caption{Random}
    \label{randomTTFT.pdf}
    \end{subfigure}
    \hfill
    \begin{subfigure}[t]{0.191\textwidth}
    \centering
    \includegraphics[width=\textwidth]{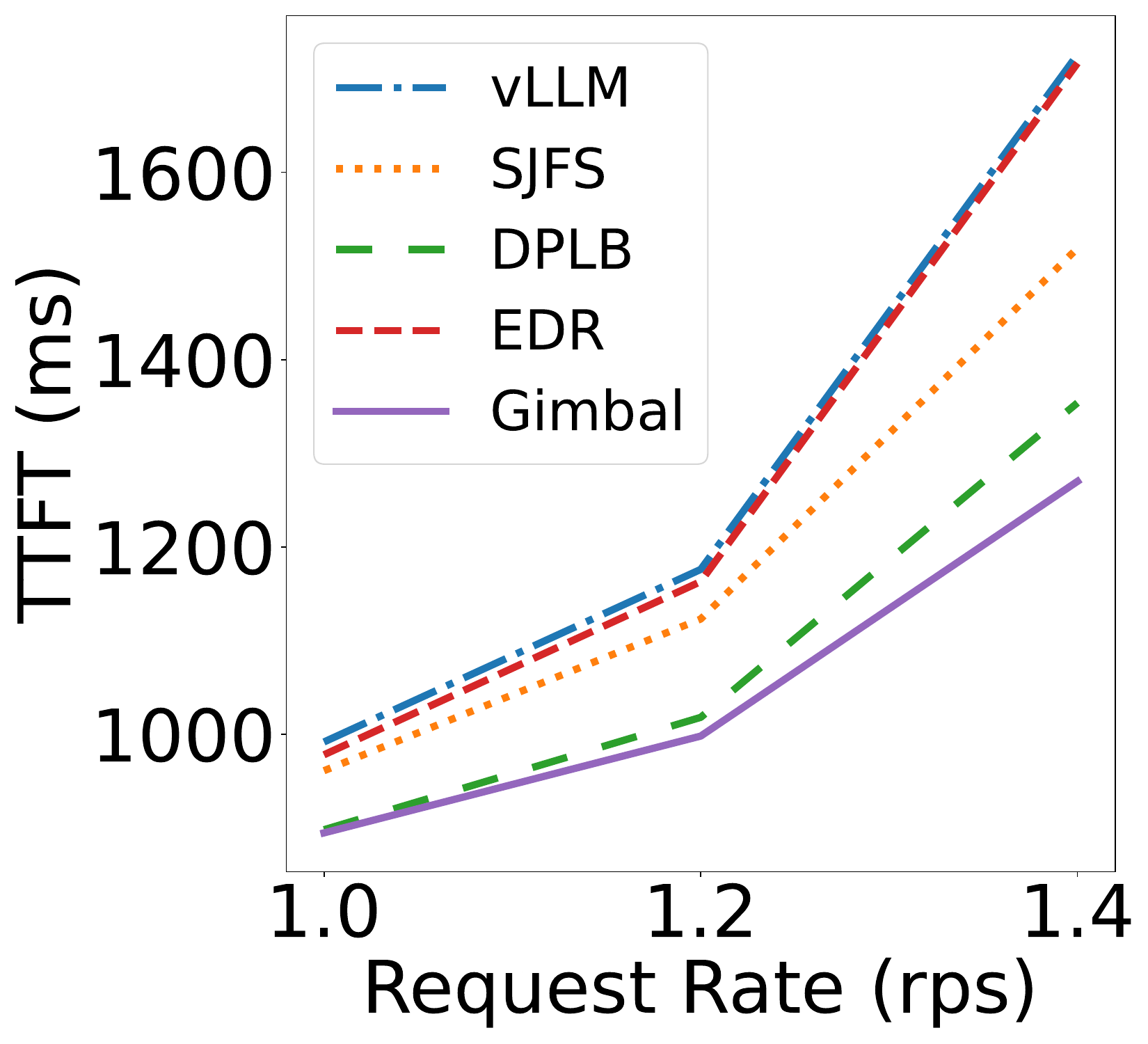} 
    \caption{Central}
    \label{centralTTFT.pdf}
    \end{subfigure}
    \hfill
    \begin{subfigure}[t]{0.191\textwidth}
    \centering
    \includegraphics[width=\textwidth]{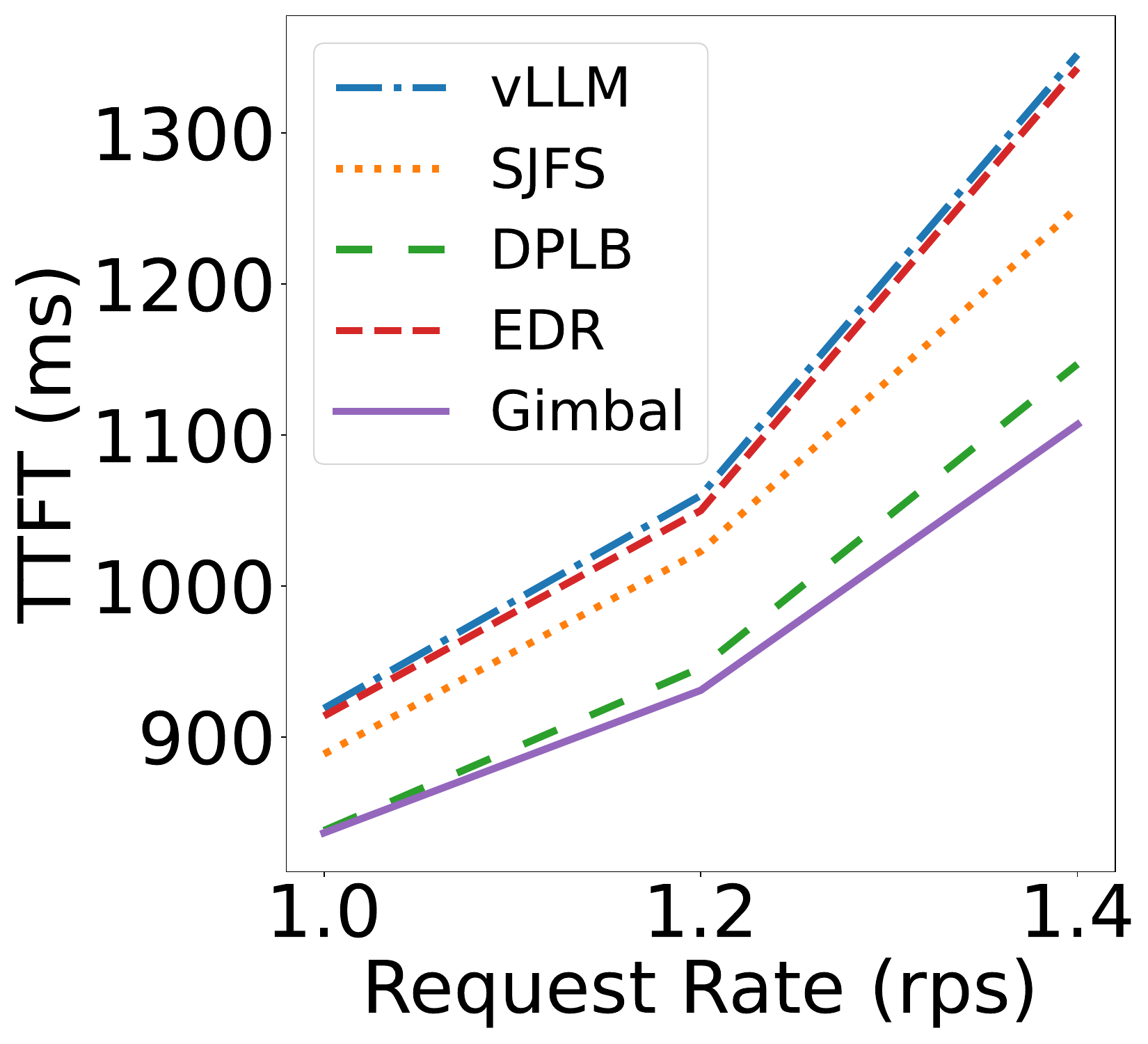} 
    \caption{Descending}
    \label{descendingTTFT.pdf}
    \end{subfigure}
    \hfill
    \begin{subfigure}[t]{0.191\textwidth}
    \centering
    \includegraphics[width=\textwidth]{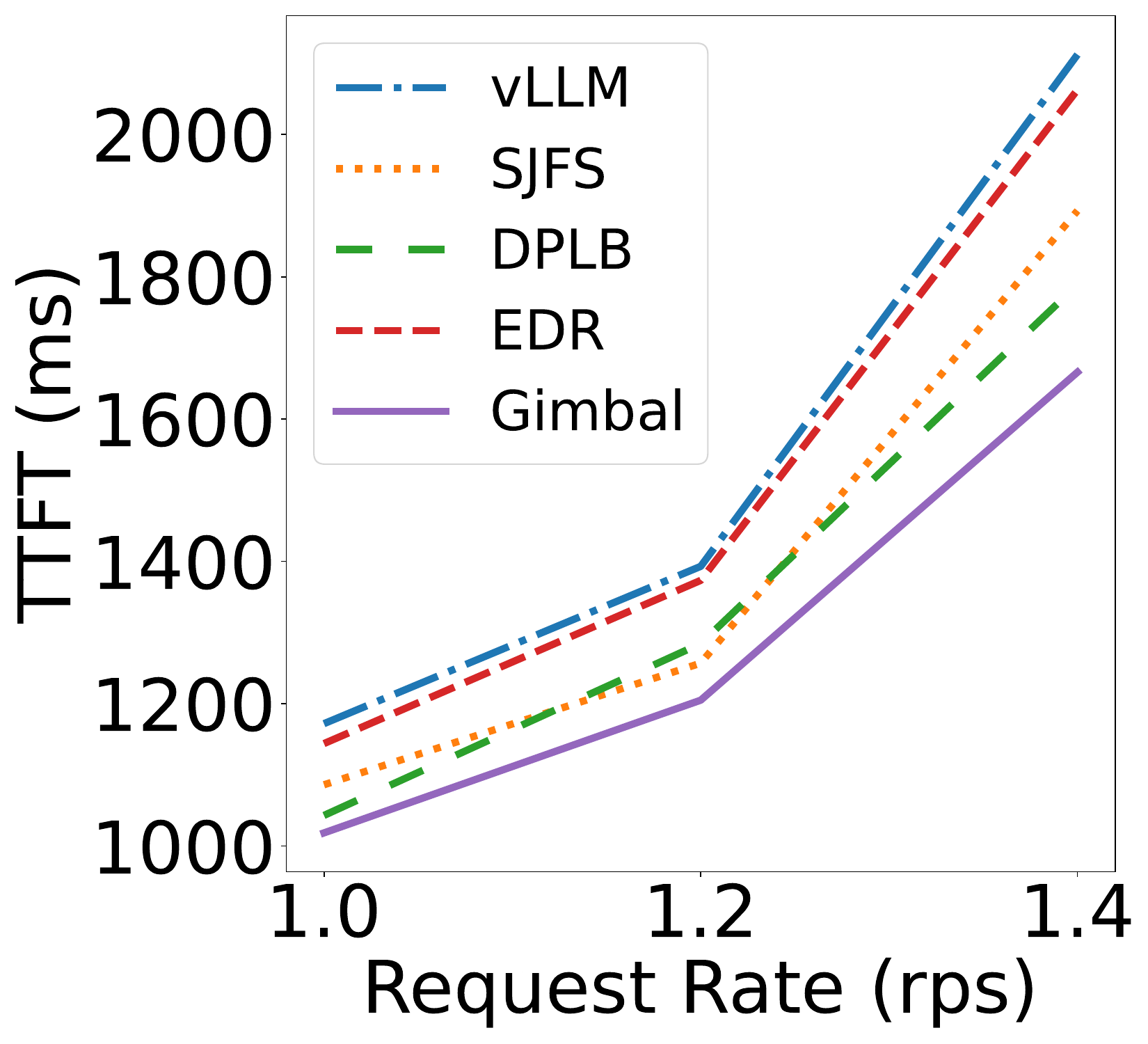} 
    \caption{Two-end}
    \label{twoendTTFT2.pdf}
    \end{subfigure}
    \hfill
    \begin{subfigure}[t]{0.191\textwidth}
    \centering
    \includegraphics[width=\textwidth]{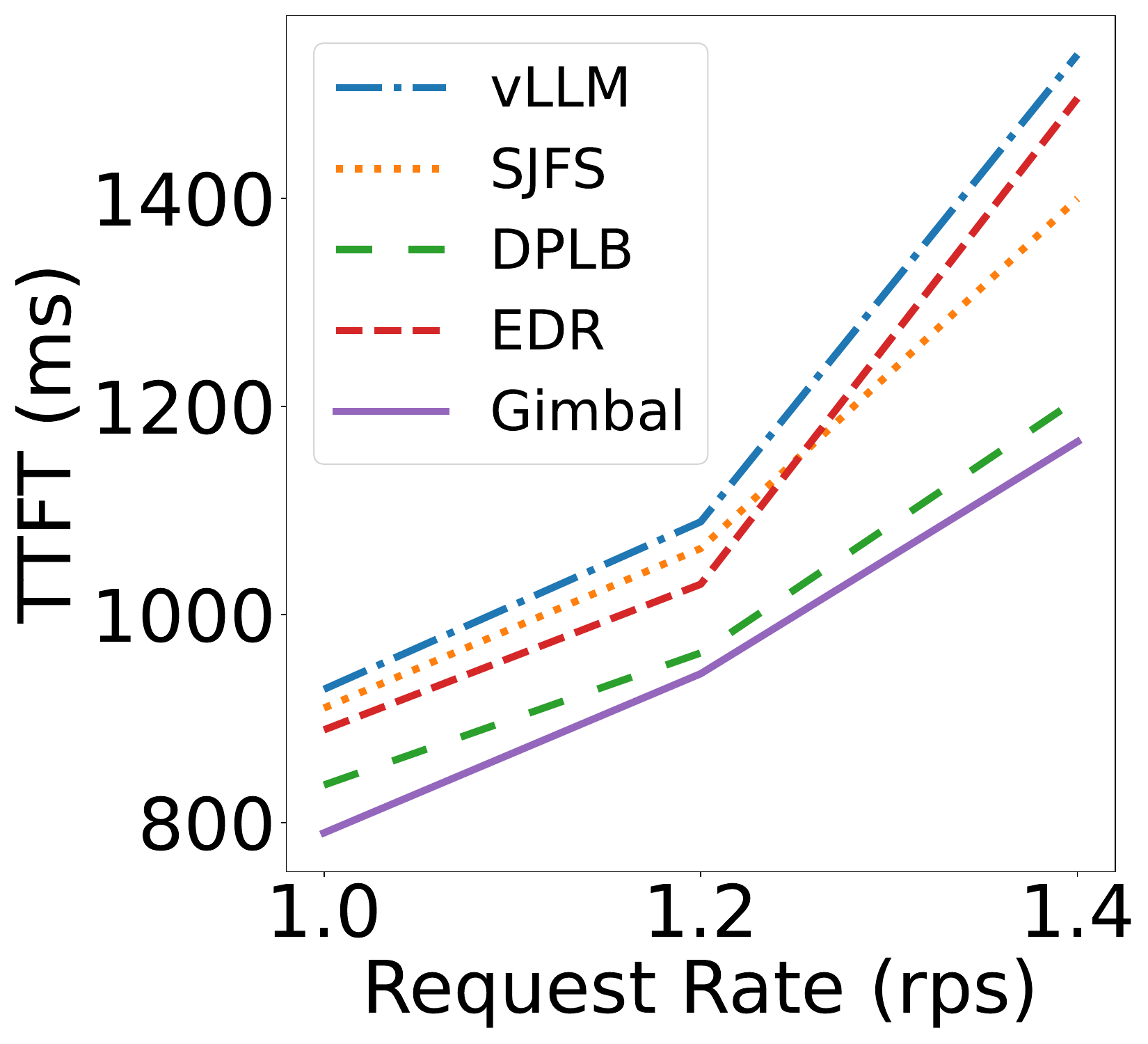} 
    \caption{Average}
    \label{averageTTFT.pdf}
    \end{subfigure}
    \caption{TTFT under different request-rate (RPS) for the five request distributions.}
    \label{fig: TTFT under different RPS}
\end{figure*}

\begin{figure}[h]
        \centering
        \includegraphics[width=0.8\columnwidth]{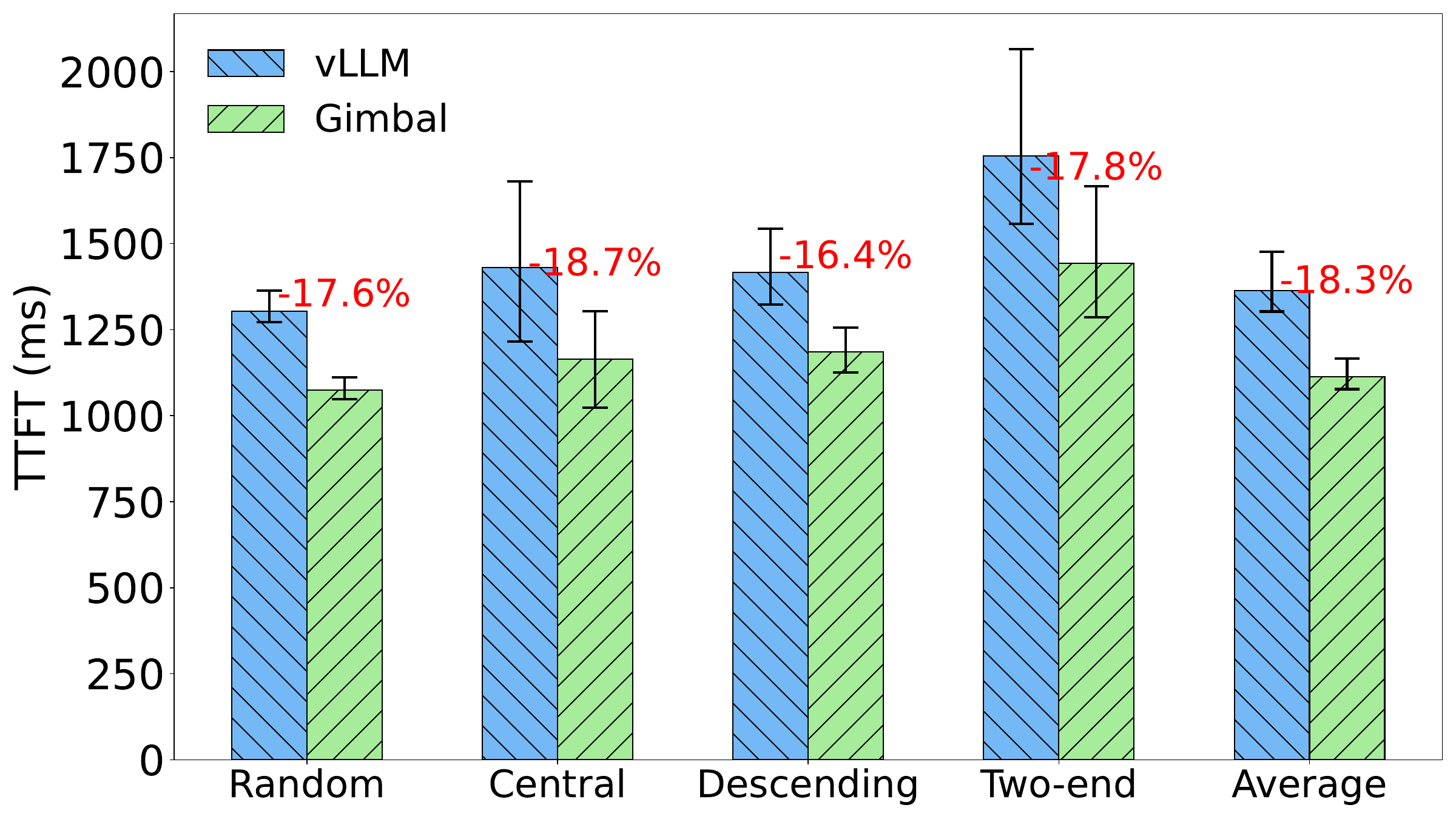}
        \caption{Average TTFT across Five Distributions at 1.4 RPS}  
        \label{fig: 1.4 RPS TTFT}
\end{figure}

Experiments are conducted on five data distributions to evaluate TTFT, with one set of 1,000 randomly selected requests per each distribution. To respect the GPU memory constraints of our testbed, the request rate is set between 1.00 and 1.40 RPS.
Figure~\ref{fig: TTFT under different RPS} presents the TTFT results for all baselines and Gimbal across five request distributions (Random, Central, Descending, Two-end, and Average) at request rates of 1.0, 1.2, and 1.4 RPS. It can be clearly observed that across all distributions, the TTFT of Gimbal consistently remains lower than that of vLLM as well as all three ablated variants SJFS, DPLB, and EDR. 
A closer examination of the ablation results suggests that DPLB accounts for the majority of the TTFT reduction, while SJFS and EDR provide additional but smaller improvements. When all three mechanisms are jointly enabled in Gimbal, their effects are complementary, resulting in the largest overall reduction.
As the system approaches saturation—e.g., in the 1.2--1.4 RPS region—the performance advantage of Gimbal becomes even more pronounced, indicating that multi-layer coordinated scheduling is particularly effective under high load.

The results also demonstrate consistent trends across all five synthetic distributions, showing that Gimbal is robust to changes in workload shape. Compared to single-module approaches, Gimbal inherits the strengths of both inter-engine load balancing and intra-engine SJF scheduling, while expert dynamic relocation further eliminates within-layer MoE imbalance, unlocking compute capacity and reducing latency.

To further reduce randomness and validate stability, we also conduct three independent repeated experiments for both vLLM and Gimbal at 1.4~RPS, where each run preserves the same workload distribution but samples different requests by using distinct random seeds. The aggregated TTFT bar chart across all five distributions, shown in Figure~\ref{fig: 1.4 RPS TTFT}, reports the mean results over three independent runs, confirming that Gimbal consistently outperforms the baseline.

Specifically, taking the mean of the three runs for each distribution, Gimbal reduces TTFT compared with vLLM by \textbf{17.6\%} on Random, \textbf{18.7\%} on Central, \textbf{16.4\%} on Descending, \textbf{17.8\%} on Two-end, and \textbf{18.3\%} on Average distribution, corresponding to \textbf{17.76\%} across all five distributions. These results verify that Gimbal can reliably reduce TTFT across different workload shapes. 

\begin{figure*}[h]
    \begin{subfigure}[t]{0.191\textwidth}
    \centering
    \includegraphics[width=\textwidth]{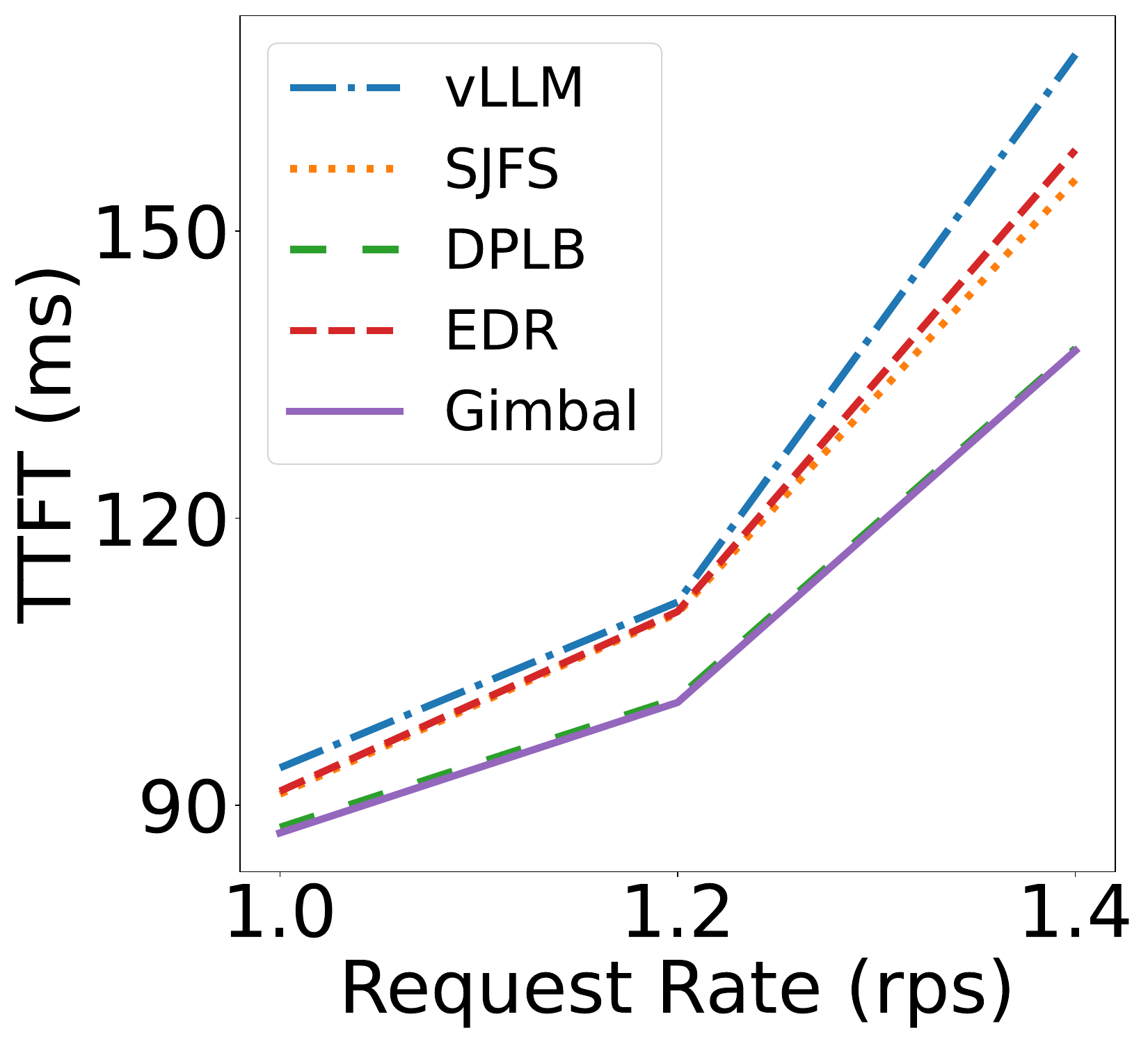} 
    \caption{Random}
    \label{randomTPOT.pdf}
    \end{subfigure}
    \hfill
    \begin{subfigure}[t]{0.191\textwidth}
    \centering
    \includegraphics[width=\textwidth]{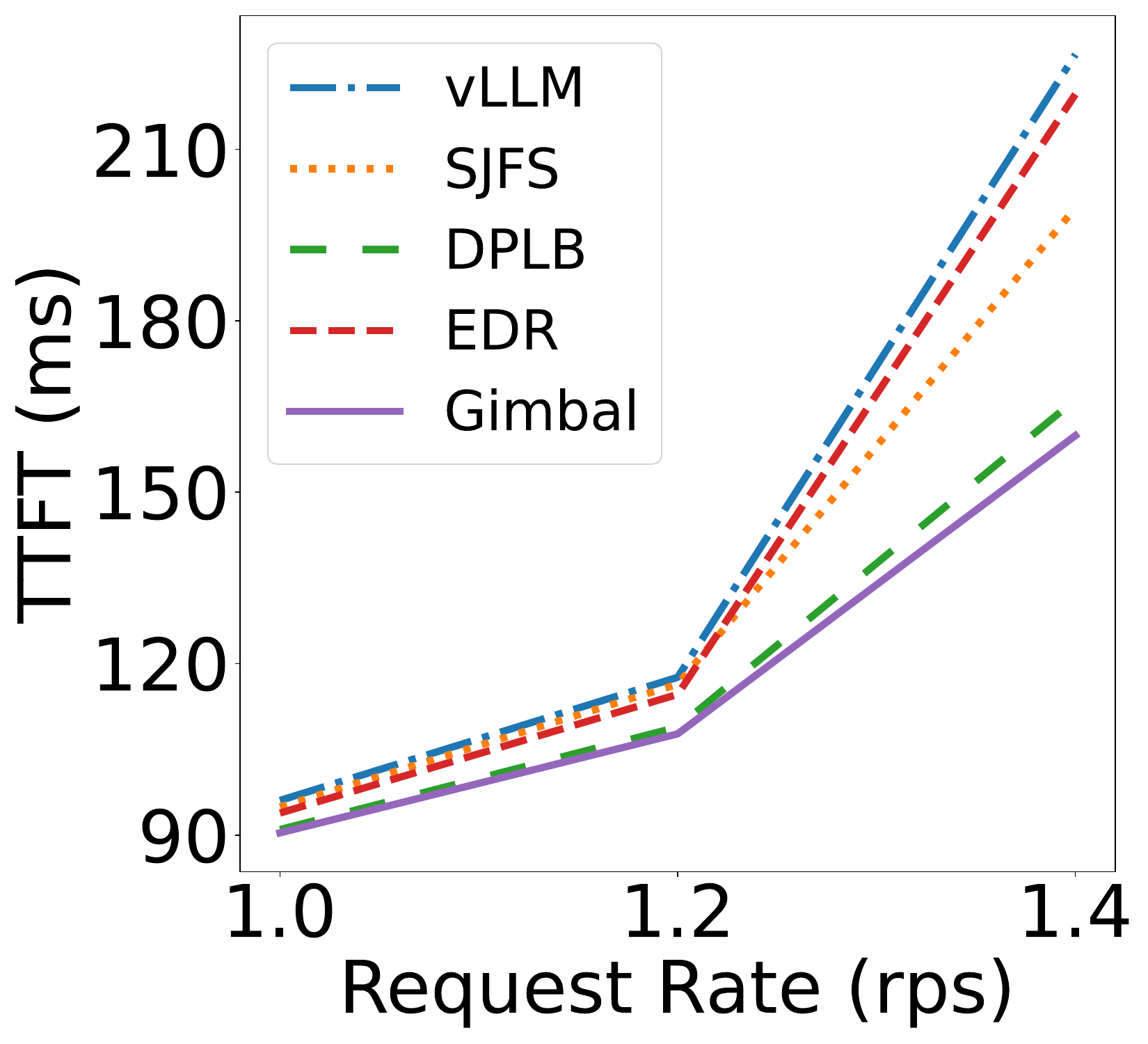} 
    \caption{Central}
    \label{centralTPOT.pdf}
    \end{subfigure}
    \hfill
    \begin{subfigure}[t]{0.191\textwidth}
    \centering
    \includegraphics[width=\textwidth]{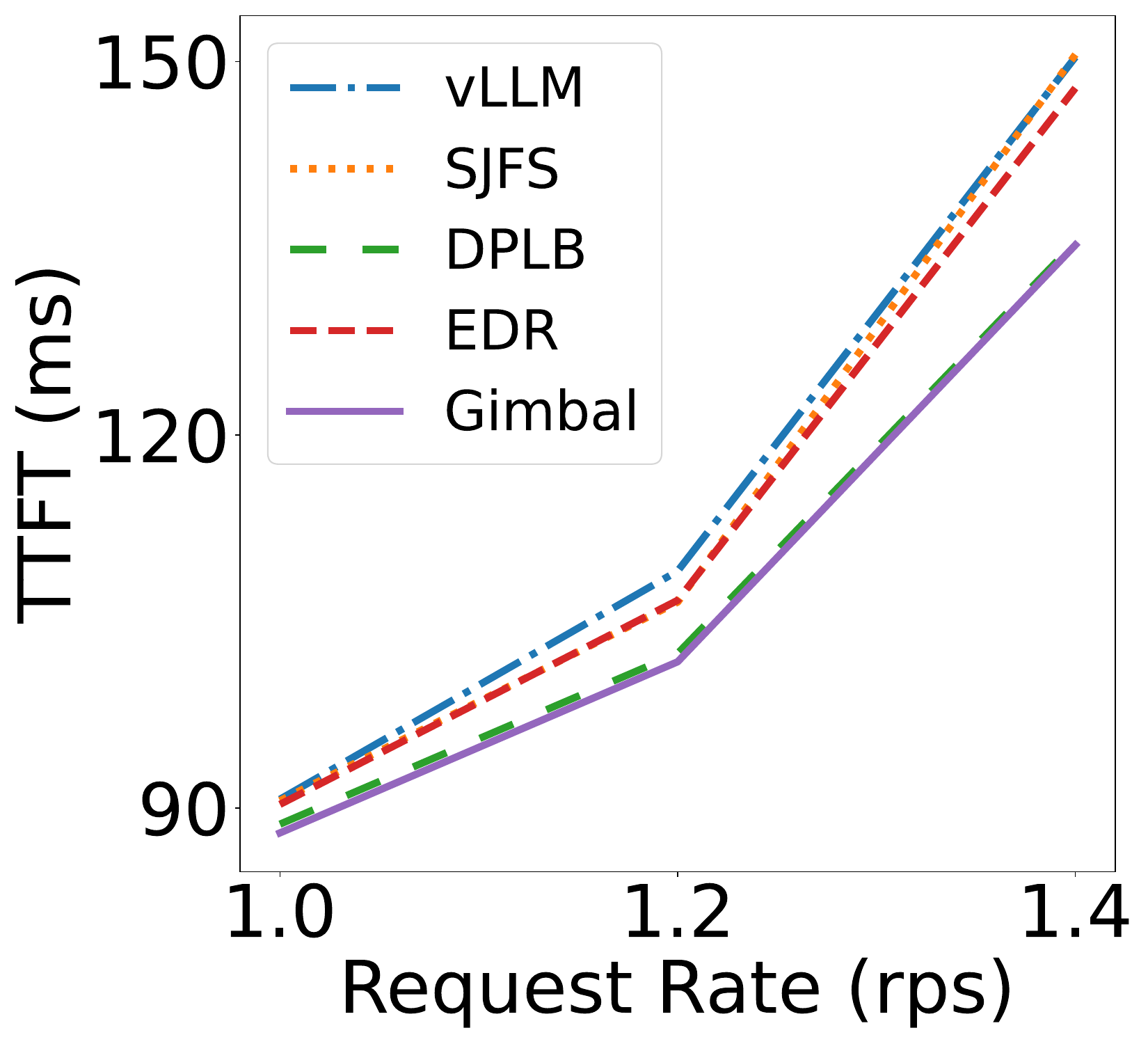} 
    \caption{Descending}
    \label{descendingTPOT.pdf}
    \end{subfigure}
    \hfill
    \begin{subfigure}[t]{0.191\textwidth}
    \centering
    \includegraphics[width=\textwidth]{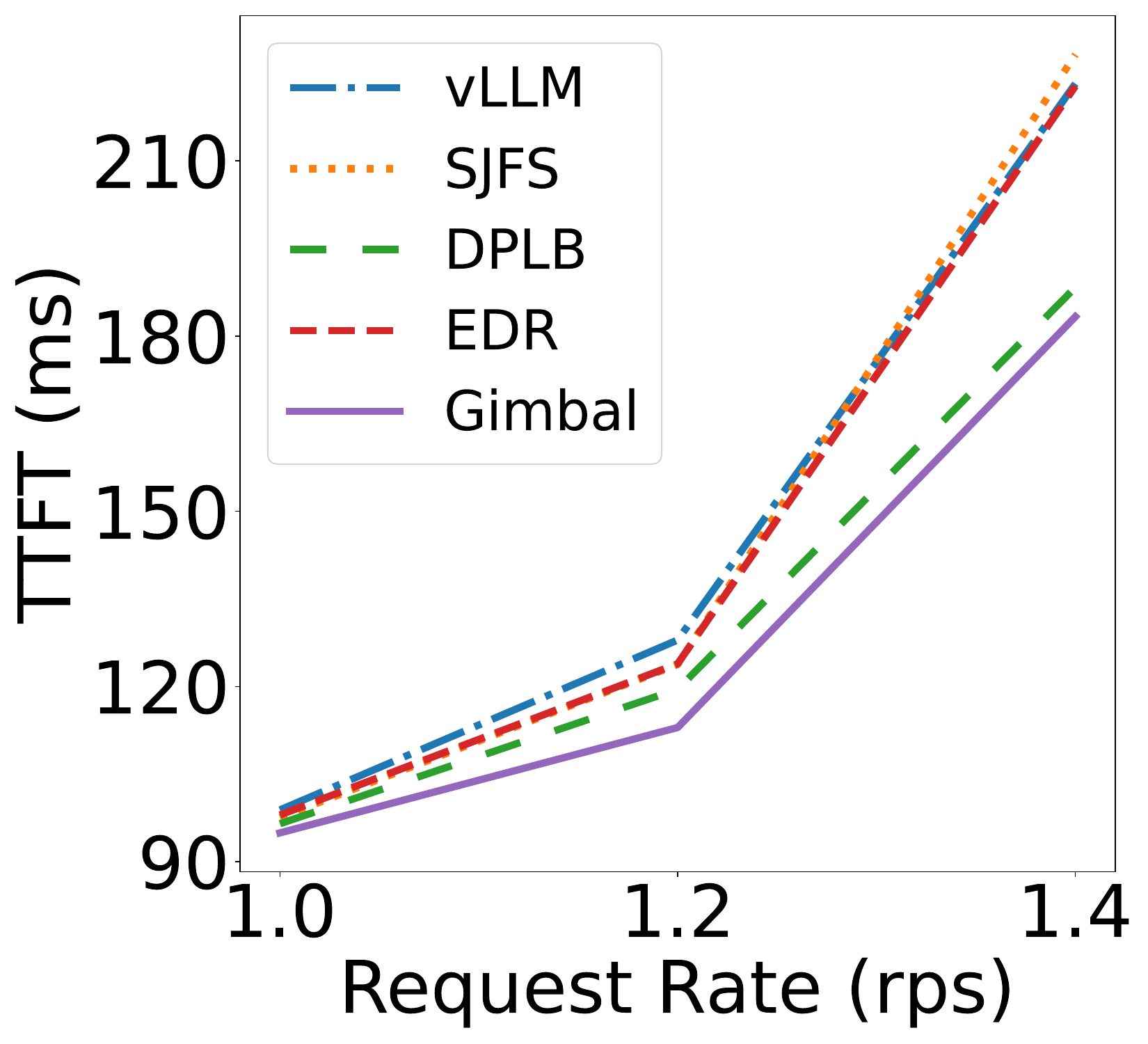} 
    \caption{Two-end}
    \label{twoendTTFT3.pdf}
    \end{subfigure}
    \hfill
    \begin{subfigure}[t]{0.191\textwidth}
    \centering
    \includegraphics[width=\textwidth]{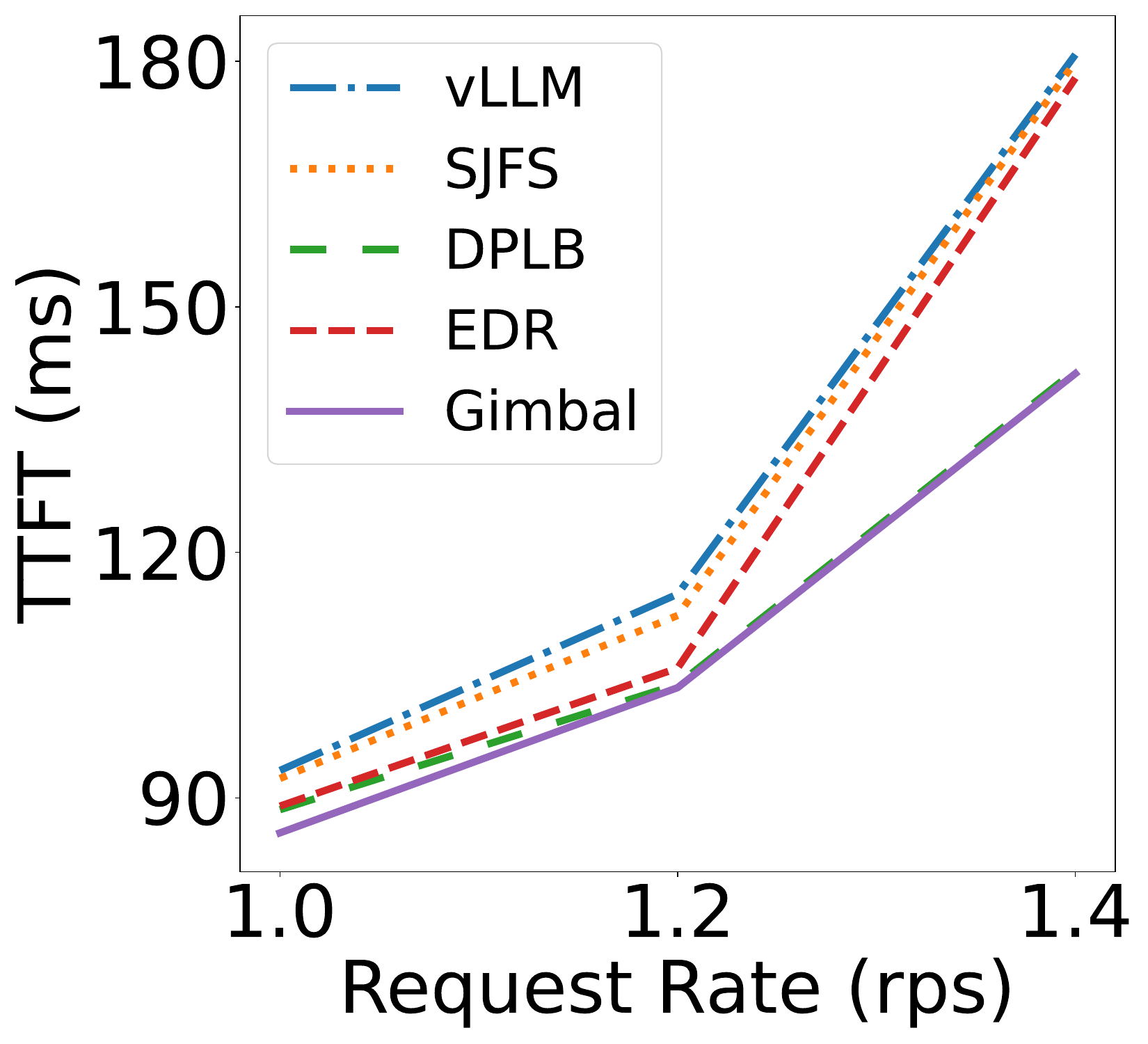} 
    \caption{Average}
    \label{averageTPOT.pdf}
    \end{subfigure}

    \caption{TPOT under different request-rate (RPS) for the five request distributions.}
    \label{fig: TPOT under different RPS}
\end{figure*}



\begin{figure}[t]
        \centering
        \includegraphics[width=0.8\columnwidth]{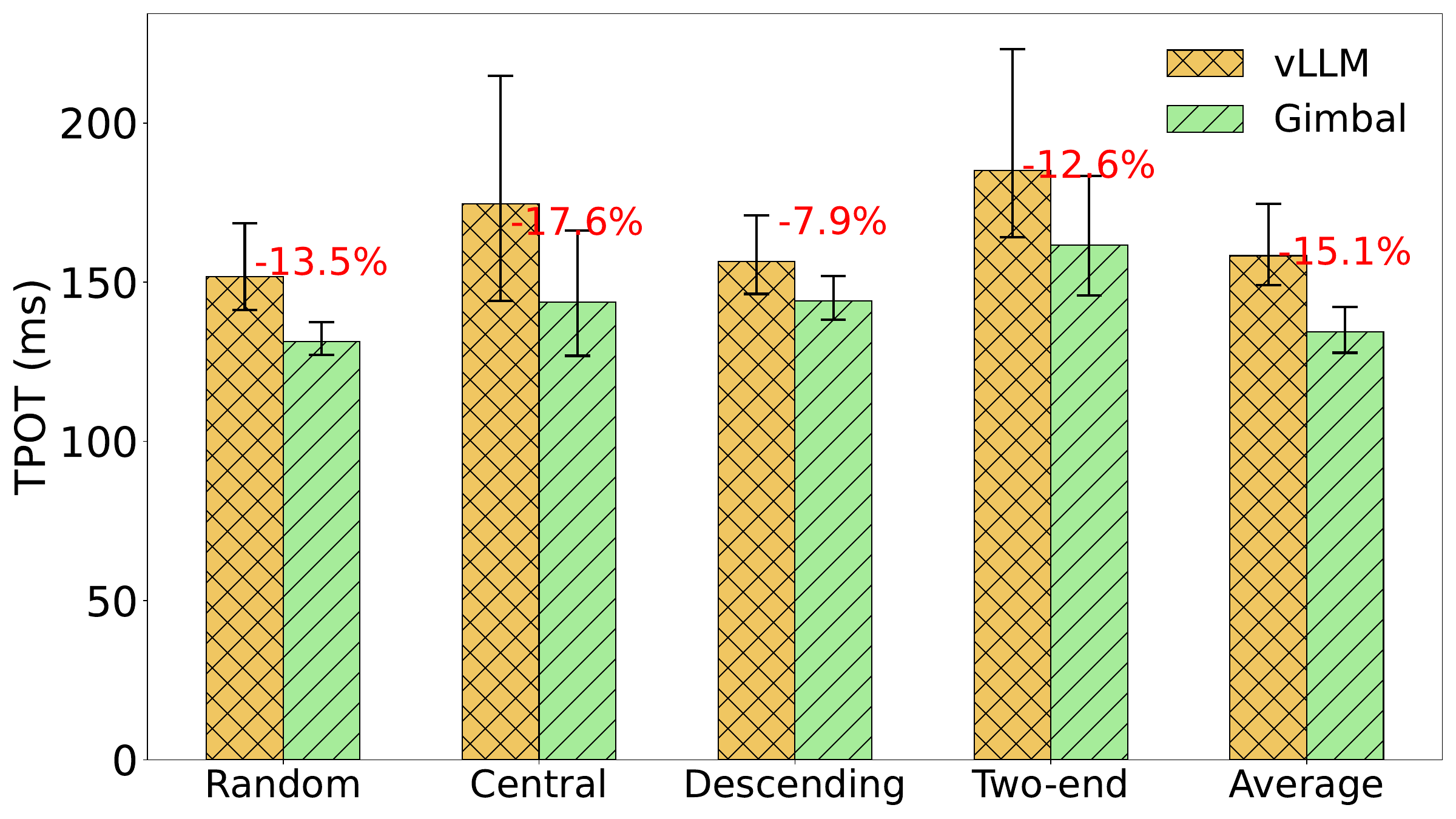}
        \caption{Average TPOT across Five Distributions at 1.4 RPS}  
        \label{fig: 1.4 RPS TPOT}
\end{figure}

\begin{figure}[h]
        \centering\includegraphics[width=0.8\columnwidth]{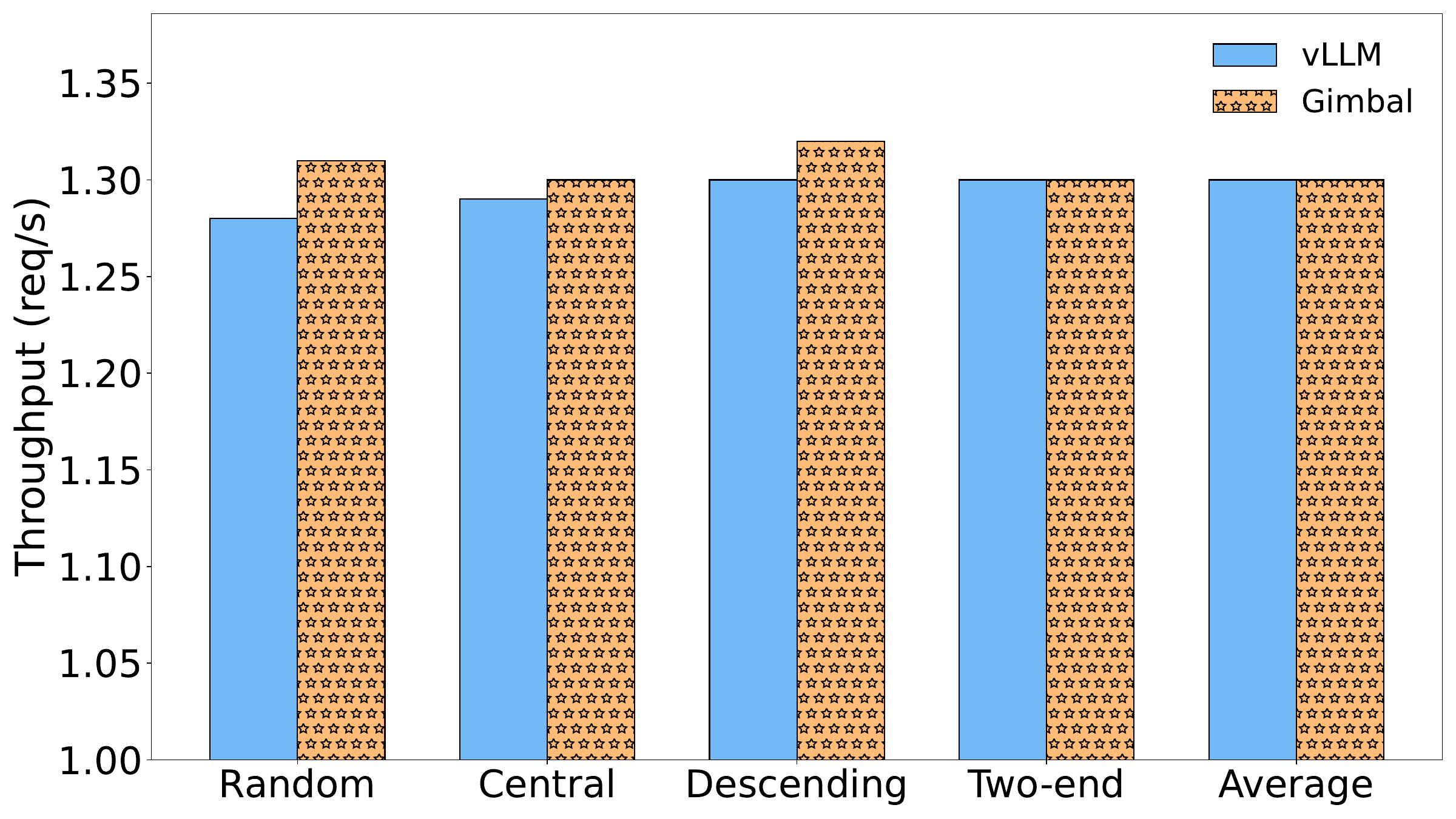}
        \caption{Average Throughput across Five Distributions at 1.4 RPS} 
        \label{fig: 1.4 Thou}
\end{figure}



\subsubsection{\textbf{TPOT Results}}

Our experiments also record the behavior of TPOT. Figure~\ref{fig: TPOT under different RPS} presents the TPOT results under five request distributions (Random, Central, Descending, Two-end, and Average). The results closely mirror those observed in the TTFT experiments: as the request Rate (RPS) load increases, the performance gaps between different variants gradually widen, while Gimbal consistently achieves lower TPOT than vLLM and all three ablated variants SJFS, DPLB, and EDR. This demonstrates that Gimbal not only optimizes TTFT but also significantly reduces the time-per-token latency, while maintaining robustness across different workload shapes.

The TPOT results from the three independent repeated runs at 1.4~RPS are shown in Figure~\ref{fig: 1.4 RPS TPOT}. Using the same aggregation method as in the TTFT analysis, we average the three runs for each distribution. Gimbal reduces TPOT compared with vLLM by: \textbf{13.5\%} on Random, \textbf{17.6\%} on Central, \textbf{7.9\%} on Descending,
\textbf{12.6\%} on Two-end, and \textbf{15.1\%} on the Average distribution. Overall, Gimbal achieves an average TPOT reduction of \textbf{13.34\%} across the five workload patterns.

Figure~\ref{fig: 1.4 Thou} reports the average throughput across the five request distributions at 1.4 RPS. As shown in the figure, Gimbal achieves throughput that is comparable to that of vLLM across all distributions, indicating that the observed latency reductions do not come at the cost of system throughput. 
Taken together, these results confirm that Gimbal can continuously reduce both TTFT and TPOT under various load conditions while maintaining comparable throughput, thereby validating the effectiveness and practicality of our approach.

\begin{figure}[h]
    \centering\includegraphics[width=0.9\columnwidth]{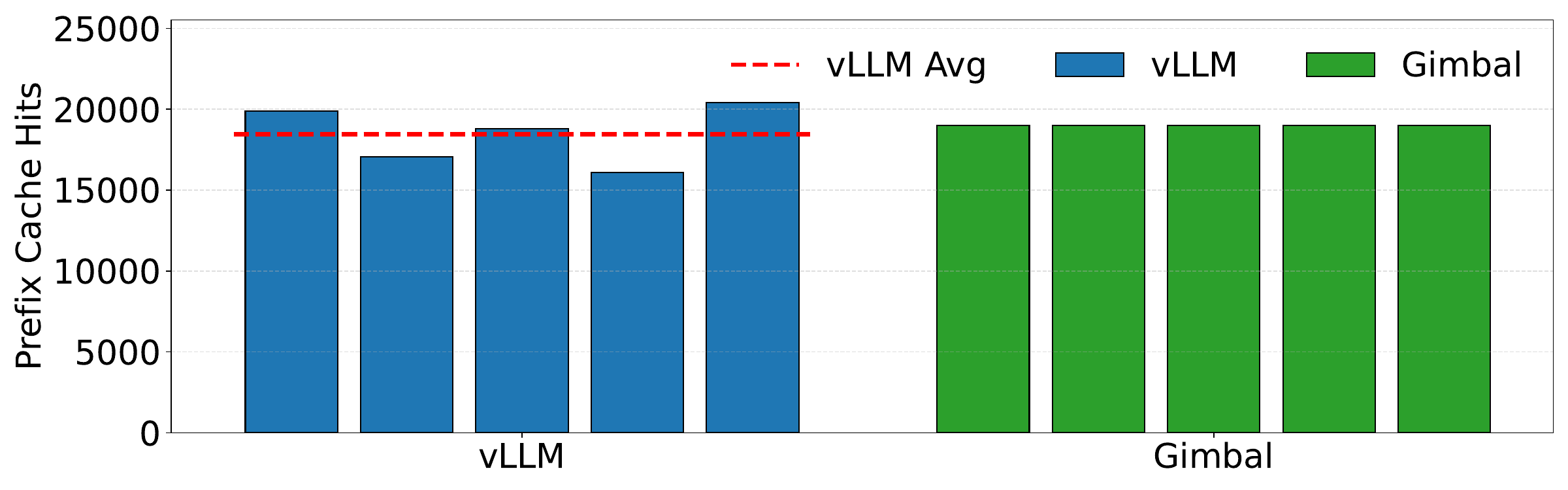}
    \caption{Prefix Cache Block Hit Count over Five Repeated Runs}  
    \label{fig: 5 rounds of prefix cache hits}
\end{figure}

\begin{figure}[h]
\centering\includegraphics[width=0.9\columnwidth]{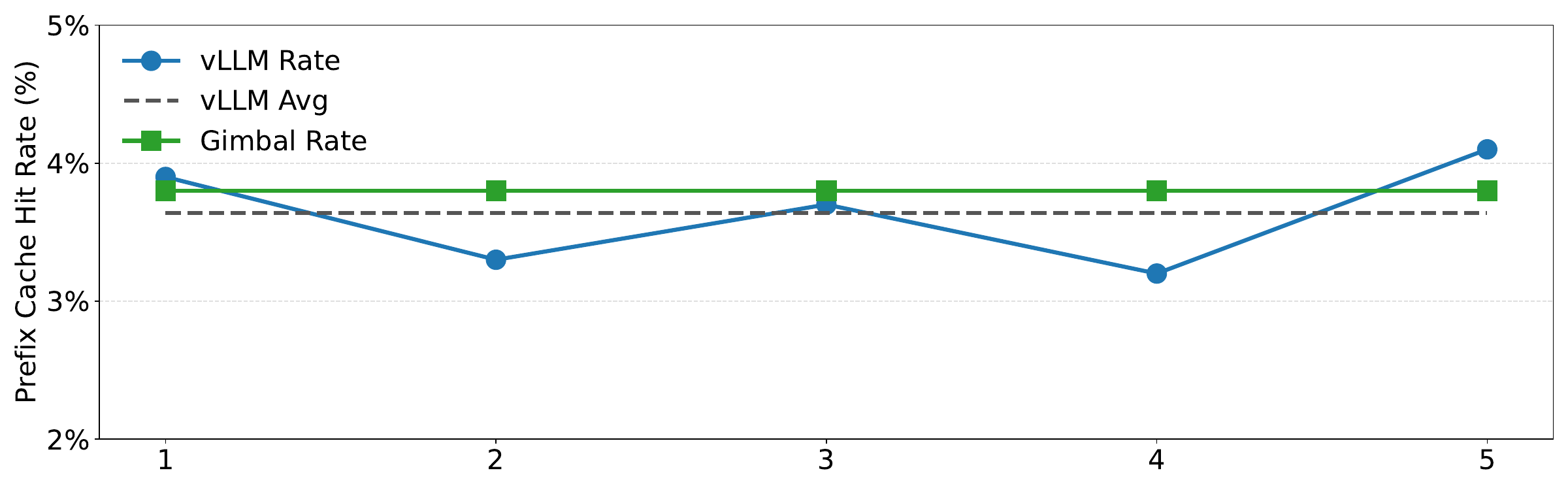}
    \caption{Global Prefix Cache Hit Rate over Five Repeated Runs}  
    \label{fig: 5 rounds of prefix cache rates}
\end{figure}

\subsubsection{\textbf{Prefix Cache Results}}
Since it is difficult to quantify how much computation is saved by a single prefix-cache hit, its direct impact on end-to-end performance is hard to measure. Therefore, we conduct a dedicated experiment to evaluate the 
user-affinity mechanism in our DP Engine LB module. We use the ShareGPT dataset~\cite{sharegpt} and run a prefix-cache–specific experiment. We run five independent experiments, each consisting of the same set of 10,000 requests for both vLLM and Gimbal.

Figure~\ref{fig: 5 rounds of prefix cache hits} reports the total number of prefix-cache hits in each round. Across all five runs, Gimbal consistently achieves higher prefix-cache hit counts than vLLM. Specifically, vLLM achieved 19,888, 17,056, 18,800, 16,096, and 20,416 hits in the five rounds, averaging 18,451 hits per round. Gimbal, on the other hand, achieved a consistent number of hits across all five runs, totaling 18,992 hits per round, representing an improvement of approximately 3\%. This behavior stems from fundamental differences in the request routing strategies of Gimbal and vLLM. By default, vLLM employs a round-robin scheduling policy, under which incoming requests—originating from both the same and different users—are distributed across engines in a largely randomized manner. 
This variability arises from multiple sources. First, although the request submission order is identical across all experimental runs, requests are dispatched to engines asynchronously, leading to non-deterministic request arrival times at each engine. As a result, the round-robin decision may observe different engine states at dispatch time across runs, causing the same request sequence to be mapped to different engines. Second, the execution time of each request is inherently variable due to differences in output length and decoding behavior, which further introduces divergence in per-engine load and queue dynamics. These execution-time variations affect the effective round-robin rotation based on in-flight requests, amplifying load imbalance across engines. Together, these factors lead to run-to-run fluctuations in request execution order and prefix-cache reuse under vLLM.
In contrast, Gimbal adopts a user-affinity scheduling strategy that consistently routes requests from the same user to the same engine. This effectively fixes the request execution order at each engine, resulting in highly stable and repeatable prefix-cache reuse. Consequently, Gimbal achieves identical prefix-cache hit counts across all runs, demonstrating more deterministic behavior and more effective exploitation of the prefill cache.

In addition to total hit counts, we also recorded the prefix-cache hit rate during each experiment. As shown in Figure~\ref{fig: 5 rounds of prefix cache rates}, vLLM achieves hit rates of 3.90\%, 3.30\%, 3.70\%, 3.20\%, and 4.10\%, averaging 3.64\%. Gimbal maintains a stable hit rate of 3.80\% across all five runs, improving the average hit rate by approximately 4.4\%.

These results demonstrate that Gimbal’s user-affinity scheduling can increase the probability of prefix-cache hits, thereby improving prefix-cache reuse, reducing redundant computation, and ultimately enhancing overall performance.

\section{Related Work}

A number of recent systems have explored scheduling for LLM serving across different layers of inference.  
At the request level, several works like Fu et al.~\cite{fu2024efficient}, Qiu et al.~\cite{qiu2024efficientinteractivellmserving}, Zhao et al.~\cite{zhao2025seallmserviceawarelatencyoptimizedresource}, Wu et al.~\cite{wu2024fastdistributedinferenceserving} propose length-aware or priority-based schedulers to mitigate Head-of-Line (HoL) blocking and improve latency, often by approximating Shortest-Job-First (SJF) or designing multi-level feedback queues. Most of these approaches rely on predicting request output lengths to approximate SJF scheduling. However, such predictions can be dataset-dependent and may not generalize well across workloads. In contrast, our approach leverages the inherent compute characteristics of inference: \emph{prefill} phases are compute-bound while \emph{decode} phases are memory-bound. We approximate SJF scheduling by directly estimating prefill workload, avoiding the pitfalls of prediction-based methods.

At the engine level, systems such as FastServe~\cite{wu2024fastdistributedinferenceserving}, SeaLLM~\cite{zhao2025seallmserviceawarelatencyoptimizedresource}, and BROS~\cite{borui2025efficientllmservinghybrid} incorporate preemption, priority scheduling, or phase-aware routing to optimize GPU utilization and service latency. Other efforts Jain et al.~\cite{jain2025performance}, and Srivatsa et al.~\cite{srivatsa2025preble} design global routers that leverage workload characteristics such as prefill/decode distinction or prefix-sharing opportunities to balance load and increase KV Cache reuse. However, these works primarily make dispatching decisions based on coarse-grained request characteristics or phase-level distinctions, without explicitly leveraging fine-grained engine-side states such as KV cache pressure, the amount of in-flight running workload, or user-level affinity. As a result, the potential of using concrete engine load signals to guide engine-level scheduling remains largely underexplored.

For expert level, QLLM~\cite{siavashi2025priority} introduces expert-level preemption with per-expert queues, while LAMPS~\cite{shahout2024fastinferenceaugmentedlarge} integrates memory-aware scheduling for API-augmented workloads. Nevertheless, these designs mainly address expert-level queuing or memory pressure in isolation, without considering cross-layer expert affinity or expert hotspot patterns that naturally arise from MoE routing behavior. As a result, they are unable to mitigate expert-level imbalance caused by skewed activation and inter-layer dependency.

Overall, while these approaches demonstrate the importance of moving beyond traditional FCFS and RR baselines, most optimizations focus on a single layer—either request scheduling, engine dispatching, or expert routing. In contrast, our work explores scheduling optimizations at multiple layers of LLM serving, including the request, engine, and expert levels. Specifically, Gimbal combines request-level SJF with, engine-level load-aware scheduling (considering prefix tokens, KV Cache usage, and user stickiness), and expert-level hotspot mitigation with dependency-aware placement, aiming to holistically improve MoE-based LLM serving efficiency.

\section{conclusions and future work}\label{sec:conclusion}
In this paper, we presented \textsc{Gimbal}, a multi-layer scheduling framework for MoE-based LLM serving that jointly optimizes request-level, engine-level, and expert-level decisions. Built on top of vLLM, \textsc{Gimbal} combines a KV- and load-aware data-parallel engine load balancer, a prefill-length–aware SJF scheduler with aging, and an expert dynamic replacement module that incorporates inter-layer expert affinity. Across more than 100 experiments on the BurstGPT and ShareGPT workloads, \textsc{Gimbal} consistently outperforms the baseline vLLM system and its single-component variants. Under five workload distributions and at high load (1.4 RPS), \textsc{Gimbal} reduces TTFT by an average of 17.76\% and TPOT by 13.34\% compared to vLLM, while also improving prefix-cache hits. These results demonstrate that coordinated multi-layer scheduling is an effective way to unlock the latent capacity of MoE-based LLM serving systems. We hope that this multi-level scheduling concept can provide insights for future LLM optimization research.

At the same time, our study has several limitations. First, all experiments were conducted on a single 2$\times$A100-80GB server with NVLink and a single MoE model. While this setup is representative of a common multi-GPU deployment, it is significantly smaller and more homogeneous than production-scale clusters. 
Second, the expert dynamic replacement module relies on an offline inter-layer affinity matrix, which incurs non-trivial data collection and runtime overhead for activation tracking and expert migration. Designing lighter-weight online affinity estimation and more cost-aware migration policies is an important direction for future work. In future work, we plan to extend Gimbal to larger and more diverse environments, support a broader range of models and workloads, and further reduce the overhead of expert-level load balancing.

\bibliographystyle{IEEEtran}
\bibliography{references}

\vspace{12pt}
\end{document}